\documentclass[twocolumn]{simpletemplate}

\graphicspath{{figs/}}

\title{ORCHID-SAGE: Fully Reproducible CALPHAD Assessment from Evidence to Phase Diagram}

\author[msen,mechanical,industrial]{Raymundo Arróyave}
\orcid{0000-0001-7548-8686}
\email{rarroyave@tamu.edu}

\affiliation[msen]{Department of Materials Science and Engineering, Texas A\&M University, USA}
\affiliation[mechanical]{J.\ Mike Walker '66 Department of Mechanical Engineering, Texas A\&M University, USA}
\affiliation[industrial]{Wm Michael Barnes '64 Department of Industrial and Systems Engineering, Texas A\&M University, USA}

\date{}

\abstract{
Knowledge of phase stability is the first requirement for materials design, and predicting the conditions under which new materials can be synthesized has become a key capability of self-driving laboratories (SDLs). For inorganic materials, notably alloys, these predictions come from thermodynamic databases built with the CALPHAD method. Building a database, however, requires expert judgment at many steps, and because these decisions are rarely recorded, others can use a database but cannot regenerate or update it. Machines can already collect the data, build a first model with machine-learned interatomic potentials and refine it against experiments. The assessment itself, the sequence of judgments that turns these parts into a reliable database, is still carried out by human experts. We argue that it can be automated reproducibly only if the expert knowledge behind an assessment is encoded in formal protocols, so that an assessment delivers a record of its construction and not only its parameters. We introduce ORCHID-SAGE, a standard that organizes this record into reproducible and automatable protocols. In this first attempt, applied to Cr--Ni, the database was regenerated from its record by a script on a second machine, by an AI agent working only through the Model Context Protocol (MCP) and by a refit from public sources only, with the same predictions. Recorded in this way, a database need no longer be fixed at publication but can be updated by AI agents as SDLs and other sources provide new evidence, while the judgment remains with people and on the record.}


\begin{document}

\section{Introduction}
\label{sec:introduction}

\begin{figure*}[tb]
    \centering
    \includegraphics[width=171mm]{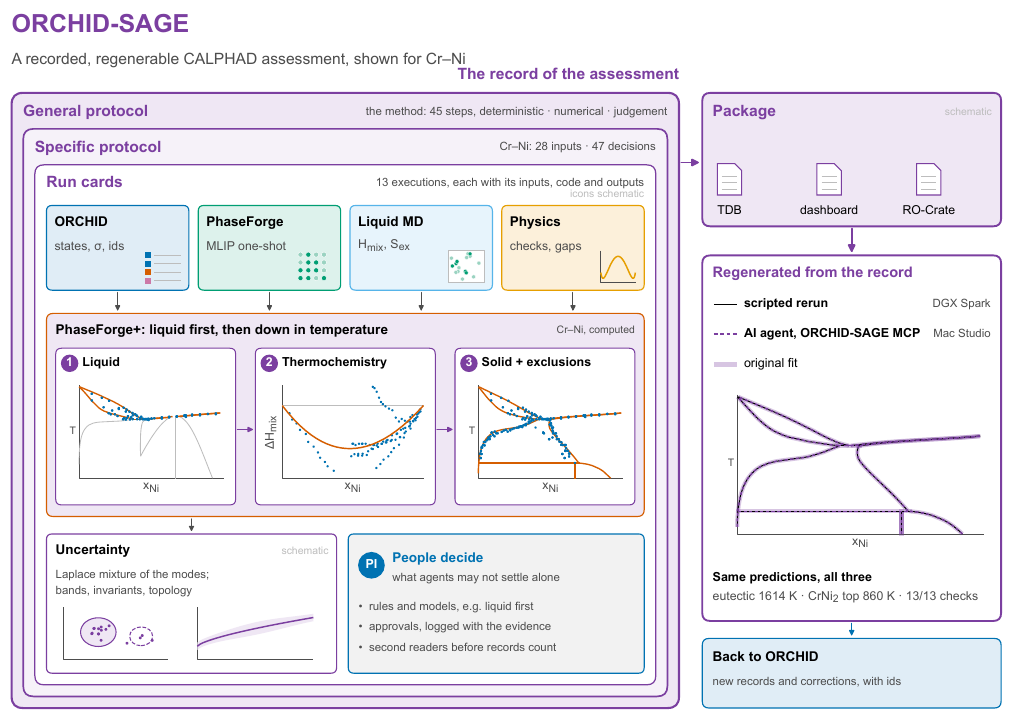}
    \caption{ORCHID-SAGE, shown for Cr--Ni. The record (general protocol, specific protocol with its decisions, run cards) wraps the assessment: ORCHID evidence, the PhaseForge one-shot model, liquid molecular dynamics, physical checks, the PhaseForge+ sequence from liquid to solid, uncertainty and the decisions made by people. The record is packaged and regenerated from itself, by a script and by an AI agent, with the same predictions. Icons are schematic. ORCHID v2026.10 and records extracted for this work; record identifiers are provided with the source data.}
    \label{fig:hero}
\end{figure*}

The CALPHAD method~\cite{lukas2007} encodes the thermodynamic properties of phases through simple, expandable models fitted to experimental and computational data. The models are then assembled into CALPHAD databases (i.e., CALPHAD model assemblages written in the standard TDB format) and fed into computational thermodynamic software~\cite{otis2017pycalphad} that, in turn, predicts a system's phase stability through constrained Gibbs free energy minimization. Over the past five decades, CALPHAD thermodynamic databases and computational thermodynamics have become central to alloy design~\cite{arroyave2022phase}.

To date, constructing CALPHAD models remains largely labor-intensive: expert assessors collect phase equilibria and thermochemical data from the literature, select a model for each phase, and adjust model parameters until the calculated phase diagram and thermodynamic properties agree with observations~\cite{lukas2007}---in most cases, the level of agreement is assigned through subjective expert opinion. Unfortunately, many of the decisions made along the way are rarely recorded in published assessments~\cite {bocklund2019espei,arroyave2026orchid}. As a result, while publicly available CALPHAD assessments can be integrated into larger databases and then used in materials discovery and development pipelines, they are extremely difficult to regenerate from scratch or update when new data become available.

Knowing the final parameters of a model fit is not sufficient to regenerate a CALPHAD assessment from scratch. At the very minimum, we also need to know (\textbf{1}) which observations were used in the fit, preferably through links to the original measurement records; (\textbf{2}) which uncertainties were assigned; (\textbf{3}) which choices were made during the fit; and (\textbf{4}) which code, software versions, and settings were used in the process.

In practice, however, a published thermodynamic model rarely reports more
than the final parameters resulting from the CALPHAD model fit, and most of the information needed to regenerate it is lost once the assessment is complete. The decisions of the assessor, such as the sequence of steps during the model optimization, the uncertainty attributed to each observation, the weight given to each dataset, or the choice between alternative models for a phase, are rarely written down and do not appear in the database file in which the assessment is distributed. In the end, only the authors can follow the path that led to the parameters, and even they---as the present author can attest---are \emph{usually unable to exactly reproduce their own fit after a few years}.

Of course, this problem is not unique to CALPHAD assessments. In bioinformatics, for example, data pipelines consist of long sequences of individual data analytics operations. In order to make these analyses reproducible, workflow managers such as Snakemake~\cite{molder2021snakemake} and Nextflow~\cite{ditommaso2017nextflow} and the CWL workflow language CWL~\cite{crusoe2022cwl} have become an integral part of the process. In the context of electronic-structure calculations, AiiDA records the complete provenance of every calculation~\cite{huber2020aiida} and repositories such as NOMAD~\cite{draxl2019nomad} distribute the inputs and outputs of millions of calculations. In many other computational fields, standard provenance descriptions, such as PROV-O~\cite{lebo2013provo}, and formats such as RO-Crate~\cite{soilandreyes2022rocrate,leo2024wrroc}, which bundle input, code, and output of a calculation, arose from the same need. Unfortunately, none of these practices has been adopted in CALPHAD model development, and, to the best of our knowledge, \emph{no thermodynamic database currently in use can be regenerated from what anyone---other than its authors---has published}.

\begin{figure*}[hbt]
    \centering
    \includegraphics[width=171mm]{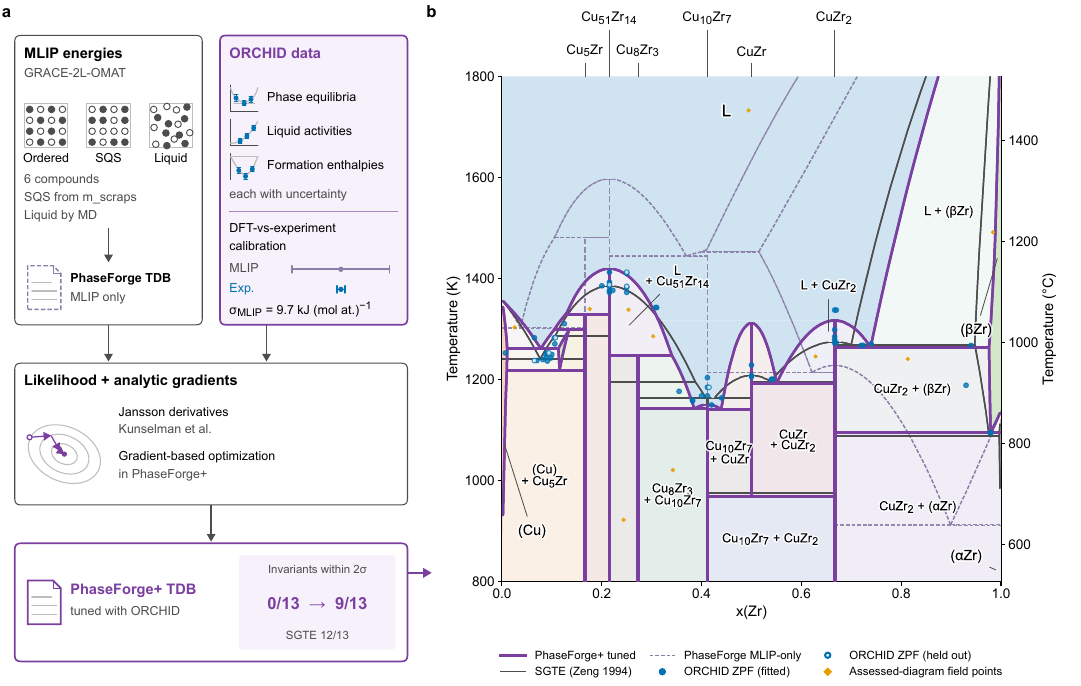}
    \caption{An example of an automated workflow for CALPHAD model refinement: \textbf{a}, PhaseForge constructs the starting description from MLIP compound, solid-solution and liquid energies on SGTE unary references, and PhaseForge+ combines these with uncertainty-weighted ORCHID observations using the gradient-capable ESPEI development branch. \textbf{b}, Refined diagram (violet), initial MLIP description (dashed) and SGTE assessment (grey), with fitted and held-out literature points and assessed assemblages. Reproduced, under CC BY 4.0 from ~\cite{arroyave2026orchid}.}
    \label{fig:workflow}
\end{figure*}

Since knowledge of phase stability is the first requirement for predictive synthesis, CALPHAD databases/models are expected to become increasingly important components in the toolkits available to self-driving laboratories (SDLs)~\cite{abolhasani2023selfdriving,szymanski2023alab}. In SDLs, CALPHAD-based predictions can help select synthesis/processing conditions to achieve a target phase constitution, exclude composition regions in which undesired phases form, and interpret the phases observed after synthesis. Phase stability predictions can thus help determine the next experiment to execute, and those experiments could, for example, update knowledge of phase stability if predictions are inaccurate. Updated phase stability observations could then, in principle, further refine the parameters of the CALPHAD database used to make the predictions in the first place. Within this new framework, a CALPHAD database used for predictive synthesis in this discovery loop \emph{cannot remain fixed and instead must evolve} as new evidence is acquired, enabling continuously updatable '\emph{living thermodynamic databases}'. However, this potentially transformative paradigm can only be realized if the process of developing CALPHAD assessments is automated and fully reproducible.

Machines can already perform several steps in constructing CALPHAD databases using open-source tools. \textbf{ESPEI}~\cite{bocklund2019espei}, for example, stores thermodynamic observations as machine-readable datasets that can be shared, combined and reused, and uses them to estimate both the model parameters and their uncertainty using Bayesian methods~\cite{honarmandi2019bayesian}. \textbf{ORCHID}~\cite{arroyave2026orchid}, which was built by AI agents deployed at scale over the literature, converts the reported data into thermodynamic states, each defined by its phases, temperature, pressure, and composition, with an estimate of its uncertainty and a link to the original measurement, so that the observations used in a fit can be identified one by one. The starting point of the fit, which in a manual assessment is usually a set of parameters estimated by the assessor, can in turn be computed with \textbf{PhaseForge}~\cite{zhu2025accelerating,zhu2025machinelearningpotentialsalloys}, which calculates the energies of the solution phases and compounds of an entire binary system with machine-learned interatomic potentials (MLIPs) and combines them with the SGTE descriptions of the pure elements into a first CALPHAD model. Such models reproduce the main features of many phase diagrams, but their features can be off by hundreds of kelvin, as in the Pt--W system~\cite{kunselman2025ptw}, and therefore must be refined against experimental data, a step that can be made more efficient and deterministic through advances such as gradient-based CALPHAD model optimization~\cite{kunselman2024analytically,kunselman2025analytical} as implemented in \textbf{PhaseForge+}~\cite{kunselman2025ptw,sariturk2025phaseforgeplus} (see \cref{fig:workflow}). The underlying engine of all these tools is, in turn, the \textbf{pycalphad}~\cite{otis2017pycalphad} software.

The data, the starting model, and the optimization can thus be handled by one of the tools described above. However, the assessment itself---that is, the sequence of judgments through which an expert turns these elements into a reliable database---has yet to be automated, and no existing tool records the decisions that connect its steps, which are precisely the ones left out of published assessments. An assessment obtained this way is therefore no easier to regenerate than one carried out by hand. As algorithms perform more of the CALPHAD assessment workflow, the number of these decisions increases, and most of them can no longer be examined from the resulting publications. A fully reproducible record of how a database was obtained then becomes the only way for others to verify, correct, or extend it, and the only way to carry over the expert knowledge behind an assessment to the machines meant to repeat it.

In this work, we address this problem with \textbf{ORCHID-SAGE} (\textbf{S}ystematic \textbf{A}ssessment of \textbf{G}ibbs \textbf{E}nergies), a standard for carrying out and recording CALPHAD assessments so others (including machines) can regenerate them. A sage is also ``a profoundly wise person'' and ``someone venerated for the possession of wisdom, judgment, and experience''~\cite{dictionarycom_sage}, and ORCHID-SAGE tries to capture the wisdom of the people who have made CALPHAD possible (see \textbf{Acknowledgments}).

Our proposed standard has three levels. At the \emph{first level}, a general protocol states the rules that apply to any assessment (for example, the conversion of the uncertainty of an observation into a weight in the objective function, the phases that must be excluded at each observed state and the criterion used to choose among the solutions found from different starting points). At the \emph{second level}, a specific protocol applies these rules to one system and records the decisions made during the assessment, the reason for each, and whether a person or an algorithm made it. At the \emph{third level}, each step of the specific protocol produces a run card that identifies the inputs, code, software versions, and outputs of the step through their hashes, and then the whole evaluation is distributed as an RO-Crate~\cite{soilandreyes2022rocrate,leo2024wrroc} together with the resulting database. Since protocols can also be read by machines, an assessment can be rerun with a script or with an AI agent that uses a Model Context Protocol server~\cite{mcp}. In both cases, the new results are compared with the recorded ones within the stated tolerances. An assessment recorded this way can therefore be checked, corrected, or extended from the protocol alone, without contacting the authors.

This standard is applied here to the Cr--Ni system, which is of considerable importance to stainless steels and Ni-base superalloys and has been assessed several times~\cite{lee1992,hao2021,kunselman2026hmc}. Measurements of the solid-state phase boundaries in this system, however, remain in conflict, making it a demanding test for an assessment in which the assessor's decisions must be recorded. The assessment starts with the evidence held in ORCHID and a PhaseForge model with the SGTE descriptions of the pure elements used as reference states. This model is then refined with PhaseForge+ in a sequence that begins with the liquid and the phases that melt, where the fewest phases compete, and continues with the thermochemical and solid-state data (\cref{fig:hero}). Throughout this sequence, we weight each observation by its uncertainty rather than by hand, keep partial tie-lines as zero-phase-fraction conditions, and exclude phases stable elsewhere in the system at states where they were not observed. We also check the model's physical consistency at each stage, using, among other quantities, the excess entropy of the liquid estimated with molecular dynamics~\cite{hong2025entropy}. We then estimate the uncertainty of the final model with Laplace approximations around the posterior modes and propagate it to the phase diagram. Finally, we show that the resulting database can be regenerated using only its specific protocol, with a script on a different machine and by an AI agent, and that predictions from the rerun agree with those of the original assessment within the stated tolerances.
\section{The ORCHID-SAGE standard}
\label{sec:methods}

Regenerating a CALPHAD assessment requires two distinct elements: a method describing how experimental and computational evidence is transformed into model parameters and a record of applying that method to a particular system. In ORCHID-SAGE, the method is specified in a general protocol, while the record consists of a small set of structured documents that conform to a common schema. Because all operations--- from selecting evidence and building the one-shot PhaseForge model to optimizing parameters and quantifying their uncertainty---must be documented using this schema, the schema largely determines which parts of an assessment can be reproduced. The description of the standard therefore begins with the documents and their schema.

\subsection{Protocol levels and the schema}
\label{sec:methods:levels}

\begin{figure}[tbp]
    \centering
    \includegraphics[width=83mm]{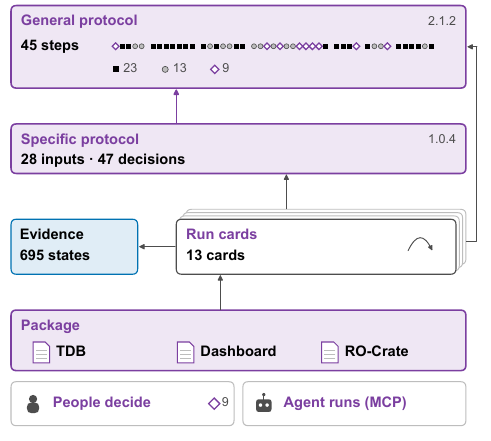}
    \caption{The ORCHID-SAGE standard, shown for Cr--Ni. The general protocol describes the method in steps of three kinds (deterministic, numerical and judgment), the specific protocol applies it to one database with its inputs, runs and decisions, and the run cards record each execution. The package delivers the database with its header, the dashboard, and the record as an RO-Crate, in which the arrows correspond to PROV relations. People decide at the judgment steps, and an agent can run the specific protocol through the ORCHID-SAGE server.}
    \label{fig:standard}
\end{figure}

An assessment consists of a sequence of three kinds of operations (deterministic calculations, numerical optimizations, and decisions requiring judgment), and regenerating its result requires a record of all three. ORCHID-SAGE maintains this record at three levels. The general protocol, at the \textbf{first level}, describes the method in a form applicable to any (binary) system. This protocol is versioned, so that the version used in this work (\texttt{orchid-calphad-fit}, version 2.1.2) can be cited unambiguously, and its written rationale is fixed through a SHA-256 hash. The specific protocol, at the \textbf{second level}, instantiates the general protocol for a single database and is linked to it through an identifier, version, and file hash. It contains the hashed inputs, the departures from the general protocol, the selected evidence, all executed runs (including failed ones), and the decisions taken, together with the person or algorithm responsible for each decision and the corresponding justification. Run cards, at the \textbf{third level}, document individual executions, including the host, interpreter, and installed packages; the exact command; the random seeds; the hashed inputs and outputs; and the optimizer trajectory. The specific protocol thus constitutes the assessment plan, while the run cards record its execution (\cref{fig:standard}).

Because the same documents are supposed to be read by humans and machines, the three levels are defined within a single schema family, \texttt{orchid-assessment-schema} (version 1.0.0-draft.2 at the time of writing), written in JSON Schema and applied to documents in a restricted subset of YAML. The organization of a protocol follows the HowTo vocabulary of schema.org~\cite{schemaorg_howto}, which divides a procedure into sections and steps. Each step includes an imperative description sufficient for execution, the supplies and tools it requires, and the results it produces. However, this vocabulary does not distinguish between steps with an outcome fully determined by their inputs and steps that depend on numerical approximations or human judgment. For this reason, the schema introduces the kind of each step, which specifies the outcome expected from a rerun: byte-identical outputs for deterministic steps, agreement within declared tolerances for numerical steps and, for judgment steps, the explicit statement of the decision rule and of the person responsible for the decision. A validator finally verifies the complete record, confirming that the documents conform to the schema, that the referenced files exist with the recorded hashes, and that no step consumes an input before it has been produced.

\begin{figure}[tbp]
    \centering
    \includegraphics[width=83mm]{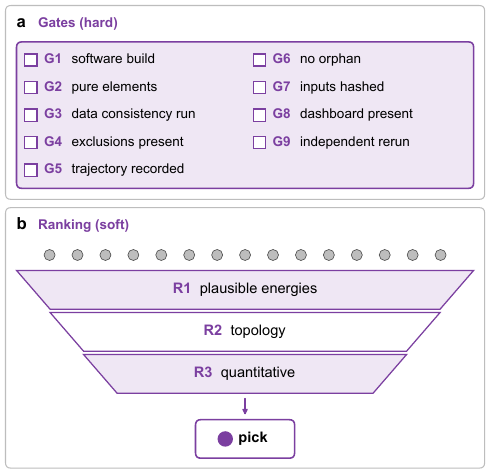}
    \caption{Gates and ranking in the general protocol. (a)~The nine gates check the integrity of the process and of the record, and an assessment that fails any of them is incomplete, however well it fits the data. (b)~Candidate fits are not gated but ranked, first by the plausibility of their energies, then by topology and finally by quantitative agreement, and the most reasonable fit is picked with its misses reported as notes. Both panels are schematic.}
    \label{fig:gates}
\end{figure}

The checks attached to the steps of a protocol serve two different purposes.~\textbf{Gates} check process integrity and record completeness. They verify, for example, that the fitting software matches the declared build, that the pure-element limits are reproduced, that excluded phases are specified at each observed state, that inputs and outputs match their recorded hashes, and that the optimizer's trajectory is stored. Before releasing a database, an independent rerun on a different machine must also agree with the original assessment. An assessment that fails any gate is considered incomplete, regardless of how well it fits the data (\cref{fig:gates}). In contrast, \textbf{fit quality} is not judged by a pass-or-fail criterion applied to a single figure of merit. Instead, we rank candidate fits first by the plausibility of the model's energy scale, then by the topology of the calculated phase diagram, and finally by quantitative agreement with the (ORCHID) data. We then select the most reasonable fit (\cref{sec:methods:sequence}).

The ranking assesses the quality of the fit, while the tolerances of the numerical steps determine whether a fit has been reproduced. Here, we note that these tolerances must account for the fact that the fitness landscape of the CALPHAD model is often nearly flat in some regions of the parameter space, and, in these directions, a rerun can cause large changes in individual parameters with minimal changes to Gibbs energies and thermodynamic observables. Thus, it is more meaningful to compare the databases' predictions than their parameters. When a rerun produces similar observables under identical protocols it is then asserted that the protocol is \emph{reproducible}.

\subsection{Evidence from ORCHID}
\label{sec:methods:orchid}

\begin{figure}[tbp]
    \centering
    \includegraphics[width=83mm]{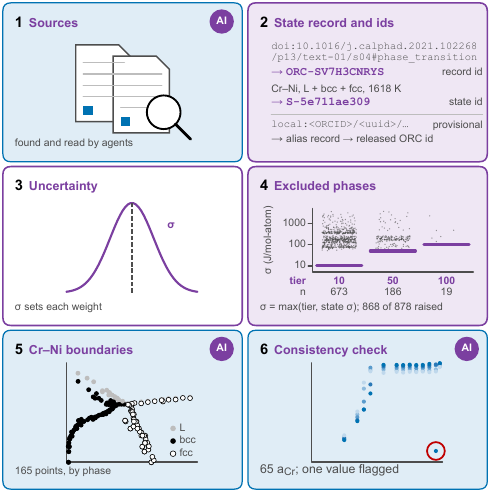}
    \caption{Evidence used in this work. (1)~Sources are found and read by AI agents. (2)~Each value is stored as a record of a fully defined state, with a record id derived from its source (here the Cr--Ni eutectic, ORC-SV7H3CNRYS) and a state id derived from the state itself. (3)~The uncertainty of each record sets its weight in the fit. (4)~Excluded phases enter with an energy scale set by their certainty tier or by the uncertainty of the state, whichever is larger. (5)~The 165 Cr--Ni phase-boundary points. (6)~The consistency check on the activities of Pugliese and Fitterer~\cite{pugliese1970}, with one inconsistent value flagged, not changed. ORCHID v2026.10 and records extracted for this work; record identifiers are provided with the source data.}
    \label{fig:orchid}
\end{figure}

Among the inputs to a CALPHAD assessment, the experimental evidence is the most difficult to trace and extract (it is dispersed in non-machine-readable forms), and for this reason, it is the first element to be fixed in the record. The evidence here comes from ORCHID~\cite{arroyave2026orchid}, which stores each (AI-extracted) value reported in the literature as an observation of a fully defined equilibrium state (\cref{fig:orchid}). For Cr--Ni (used in this work as an example), ORCHID's public release contains 42 accepted values, mostly corresponding to invariant reactions, phase transitions, and calculated enthalpies, but no experimental thermochemical data for the liquid and solid solutions. Therefore, AI agents were used to follow the citation chains of the existing assessments, retrieve the articles that were not available locally, and extract, using the ORCHID schema, the data compiled by Hao et al.~\cite{hao2021} as well as the Cr activities measured by Pugliese and Fitterer~\cite{pugliese1970}. In each case, the value was traced to the publication in which it was first reported.

For use in a reproducible assessment, these observations must be independently identifiable. ORCHID records carry identifiers derived from the locator of the value in the source (for example, the DOI of the paper together with the table and row in which the value appears). The equilibrium states observed in these records receive unique identifiers of their own, obtained as a hash of a canonical description of the state that includes the system, the phases, the pressure and the coordinates, so that \emph{two people describing the same state arrive at the same identifier}. Data not yet released by ORCHID, such as the new extractions of this work, are assigned identifiers in a provisional namespace constructed from the ORCID of the contributor and an identifier of the dataset. Upon publication, these provisional identifiers are linked to the released ones through alias records, and they are never rewritten or reused. Each state in the evidence of an assessment thus carries either its ORCHID identifiers and the release in which they appear, or the reason why this link is still missing, together with a reference to the intake through which it will be resolved.

When performing a CALPHAD model fit, different pieces of evidence must be provided with a statistical weight~\cite{bocklund2019espei} that is in turn used to compute to the likelihood (or model error) of the CALPHAD fit. Rather than assigning these uncertainties 'by hand', we take this uncertainty from the source when reported there. Otherwise, we use the calibration of the corresponding measurement method in ORCHID.

We note that an 'observation' contains more information than what has been directly observed or inferred, e.g., the location of a phase boundary, the phases present at a given composition and temperature, a thermo-chemical measurement. In fact, what has \emph{not been observed} is also valid (and, in some cases even stronger) evidence. More specifically, if an expert human assessor is fairly certain of the phases likely to be present in the equilibrium state under given thermodynamic conditions, then they can also be certain of the phases that \textbf{must not be} part of that state.

For this reason, the ORCHID standard requires the definition of a list of excluded phases at all observed states, including single-phase states, and a dataset without this list is considered incomplete, although the absence of a so-excluded phase is never claimed to be proven. The strength of an exclusion depends on the \emph{thermodynamic distance} between the state and the excluded phase field, measured relative to the state's uncertainty in temperature and composition, and is expressed through certainty tiers that set the energy scale of the corresponding term in the objective function. This energy scale can be understood to be a minimal distance away from the equilibrium hyper-plane (or convex hull) at a given thermodynamic condition. A scale of \qty{10}{\J\per\mole} of atoms is assigned when the state lies more than three uncertainties away from any observed field or boundary of the phase and no source reports the phase there, a scale of \qty{50}{\J\per\mole} when the state lies closer or the exclusion rests on a single source, and a scale of \qty{100}{\J\per\mole} when the state is adjacent to the boundary of the phase or the exclusion is assumed from a diagram or contested.

This three-tiered scale reflects the degree of 'thermodynamic certainty' that can be assigned to a phase exclusion, with higher tiers corresponding to larger uncertainty. The assigned tier, however, sets only a lower bound, since the scale used is the larger of the tier and the state's uncertainty, converted to energy through the entropy change across the boundary. In this way, an exclusion can never be held more tightly than the measurement on which it rests allows, and it is the evidence, not the default settings of the fitting software (ESPEI in this case~\cite{bocklund2019espei}), that determines how strongly the phase diagram's topology is held in place.

Before entering the fit, the data are also checked for internal consistency. For Cr--Ni, the AI agents carried out this check and found that the activity of 0.112 printed by Pugliese and Fitterer~\cite{pugliese1970} at 95~at.\% Cr and \qty{1448}{\K} cannot be reconciled with the cell potential reported in the same row of their table, which is almost identical to those of the neighboring alloys. The same agents also found that one of the phase-boundary datasets contains six repeated rows. We note that neither finding changes the data, since the suspect activity is kept with a flag and enters the fit with its uncertainty, while the repeated rows are counted only once.

Once checked, the evidence read by the fit is frozen in a versioned evidence bundle. This bundle contains the datasets and the files of excluded phases exactly as they are read by the objective function, the observations derived from the one-shot model constructed using PhaseForge~\cite{zhu2025accelerating} and the inputs required for a rerun, together with a record of the observed states, their certainty tiers and their links to ORCHID. A manifest lists the SHA-256 hash of each of these files as well as a hash of the bundle as a whole. Since a bundle is never overwritten, any change in the evidence, including correcting an error, produces a new version of the bundle, which can then be cited by a new version of the specific protocol. Only derived facts enter the bundle, whereas restricted sources are referred to only through their hashes and are not redistributed. A rerun thus requires only the bundle, while the original sources are needed only to rebuild the bundle from the beginning.

\subsection{The one-shot model}
\label{sec:methods:phaseforge}

After specifying the evidence, the next element of the record is the model used as the starting point for the fit. In this work, this starting point is the one-shot model obtained using PhaseForge~\cite{zhu2025accelerating,zhu2025machinelearningpotentialsalloys}, constructed entirely from energetics calculated with a MLIP. The model construction procedure is part of the general protocol, so it cannot be changed between assessments without a recorded decision. In the one-shot model, the Gibbs energies of the end-members are taken from the SGTE descriptions of the pure elements~\cite{dinsdale1991sgte}, while the potential is used to calculate the mixing energy of each phase relative to the pure elements on the same lattice. This makes the model compatible with existing thermodynamic databases.

\begin{figure}[tbp]
    \centering
    \includegraphics[width=83mm]{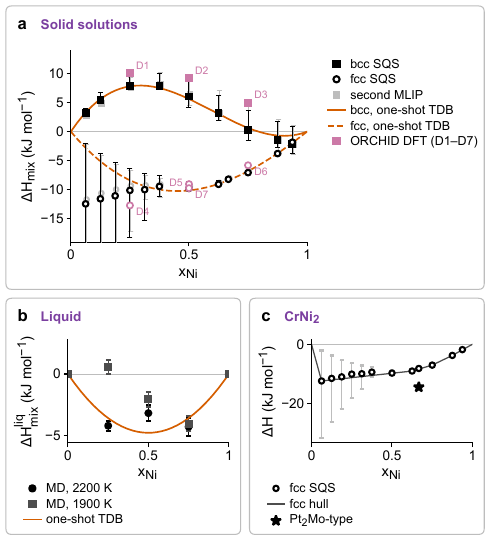}
    \caption{Energetics of the one-shot model of Cr--Ni, in \unit{\kilo\joule\per\mole} of atoms. (a)~Mixing enthalpies of the bcc (squares) and fcc (circles) SQS, calculated with GRACE-2L-SMAX-OMAT-large (black) and GRACE-2L-OMAT (grey), with the interval between the unrelaxed and fully relaxed energies as error bars, the Redlich--Kister terms of the one-shot model (orange) and the DFT records of ORCHID (pink)~\cite{liu2016alcocrni,bracq2021enthalpy}. (b)~Mixing enthalpy of the liquid from molecular dynamics at \qty{2200}{\K} and \qty{1900}{\K}, with the single term of the one-shot model. (c)~CrNi$_2$ against the fcc SQS and their convex hull. ORCHID v2026.10; record identifiers are provided with the source data.}
    \label{fig:energetics}
\end{figure}

For the bcc and fcc solid solutions, we obtained the mixing energies from 128-atom special quasirandom structures (SQS)~\cite{zunger1990special}, generated with m\_scraps~\cite{arroyave2026mscraps,singh2021scraps} using two different seeds at each composition, so the spread between the two seeds provides an estimate of the configurational uncertainty. We evaluated these structures with the GRACE-2L-SMAX-OMAT-large potential~\cite{bochkarev2024grace}, using GRACE-2L-OMAT as a cross-check. A random arrangement of atoms on a lattice that is intrinsically unstable at a given composition tends to transform during relaxation, and the energy obtained this way no longer corresponds to the phase being modeled. To detect such transformations, we relaxed each structure in stages: first in volume, then in volume and ionic positions, and finally in cell shape, and checked the lattice after each stage. The energy after the relaxation of volume and ions is the value used in the model, while the interval between the unrelaxed and the fully relaxed energies defines its uncertainty. In Cr--Ni, the bcc structures with $x_{\mathrm{Ni}} \geq 0.5$ and the fcc structures with $x_{\mathrm{Ni}} \leq 0.31$ transformed even at this intermediate stage. Instead of being discarded, these structures enter the model with the midpoint of the interval as their value and half of the interval as their uncertainty, reflecting the fact that their energy is only known to lie between these two limits (\cref{fig:energetics}a).

The mixing energies were then fitted with Redlich--Kister polynomials, with the interaction parameters defined in alphabetical order of the elements, as is the convention in CALPHAD databases. Since there is no evidence of chemical ordering in the fcc or bcc solid solutions of this system, these phases are described as disordered solutions (\texttt{FCC\_A1}, or \texttt{BCC\_A2} in Strukturbericht designation ), while CrNi$_2$, which has the Pt$_2$Mo structure, is described as a stoichiometric compound with a formation energy calculated with the same potential (\cref{fig:energetics}c).

When the one-shot model is later refined against experimental data, its mixing and formation energies enter the fit as observations in their own right, so the refined model cannot depart from them without a cost. Each observation is assigned an uncertainty of \qty{9.7}{\kilo\joule\per\mole} of atoms, corresponding to the typical difference between energies calculated with the density functional theory on which the potential is trained and calorimetric measurements~\cite{kirklin2015oqmd,arroyave2026orchid}.

Unlike the solid phases, the liquid cannot be represented by static structures, so we calculated its mixing energy through molecular dynamics simulations with the same potential. We performed simulations with 108 atoms in five compositions, including pure liquids as references, and at two temperatures: \qty{2200}{\K}, just above the melting point of Cr, and \qty{1900}{\K}. In all cases, production runs of \qty{100}{\ps} followed the melting and equilibration of the cell, and the liquid character of the final configuration was confirmed through the mean-square displacement of the atoms. We assessed the effect of cell size with an additional simulation of 864 atoms at $x_{\mathrm{Ni}} = 0.5$ and \qty{2200}{\K} (\cref{sec:methods:entropy}). The enthalpy of mixing at each temperature corresponds to the average total energy, including the $pV$ term, relative to that of the pure liquids (\cref{fig:energetics}b), and these enthalpies were fitted with a single Redlich--Kister parameter, the temperature dependence of which is determined by the excess entropy of the liquid (\cref{sec:methods:entropy}). This description of the liquid, together with the solid-phase energies, completes the one-shot database. The database and the phase diagram calculated with it in pycalphad~\cite{otis2017pycalphad} thus constitute the first result of the assessment, and the reference against which the subsequent refinement is measured.

\subsection{Liquid entropy}
\label{sec:methods:entropy}

\begin{figure}[tbp]
    \centering
    \includegraphics[width=83mm]{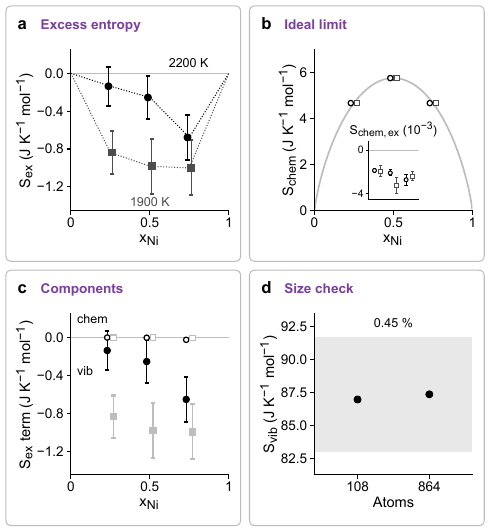}
    \caption{Excess entropy of the Cr--Ni liquid from molecular dynamics, in \unit{\J\per\K\per\mole} of atoms, at \qty{2200}{\K} (circles) and \qty{1900}{\K} (squares), with 108 atoms and \qty{100}{\ps} of production. (a)~Total excess entropy; the error bars do not include the error of the potential. (b)~Ideal-limit check: with the species labels permuted at random, the pair estimator returns the ideal entropy of mixing (grey). (c)~Vibrational (filled) and chemical (open) excess terms. (d)~Cell-size check at Cr$_{50}$Ni$_{50}$ and \qty{2200}{\K}: 108 and 864 atoms differ by 0.45\%, within the $\pm 5\%$ criterion. The pure-Cr reference at \qty{1900}{\K} crystallized after \qty{82}{\ps} and is evaluated over its first \qty{80}{\ps}.}
    \label{fig:entropy}
\end{figure}

Because the mixing enthalpies at two temperatures are insufficient to determine the temperature dependence of the liquid parameters, we also estimated the excess entropy of the liquid from molecular dynamics simulations, following the method proposed by Hong and Liu~\cite{hong2025entropy}. In this method, the entropy of the liquid is separated into two contributions, a vibrational one, obtained from the spectrum of the velocity autocorrelation function, and a configurational one, accounting for the chemical mixing of the two species. In its original form, however, the chemical-mixing term does not reduce to the ideal entropy of mixing for a random distribution of the species, so we replaced it with the pair (quasi-chemical) entropy of the species on the neighbor graph of the simulations, which recovers the ideal limit exactly, as shown in \cref{app:entropy}. Before any excess entropy was calculated, the ideal limit was verified on the simulations themselves, with the species labels assigned at random (\cref{fig:entropy}b). Because the corrected estimator has not yet been independently reviewed, we treat the excess entropies obtained with it as provisional throughout this work.

The excess entropy of the liquid at each composition and temperature corresponds to the entropy of the mixture minus the entropies of the pure liquids at the same temperature and the ideal entropy of mixing. For Cr--Ni, this quantity is small and negative, ranging between about $-0.1$ and $-1.0$~\unit{\J\per\K\per\mole} (\cref{fig:entropy}a), and it is dominated by the vibrational contribution, with the chemical contribution being two orders of magnitude smaller (\cref{fig:entropy}c). The excess entropies enter the fit as observations, with an uncertainty that combines the statistical error of the simulations with the error of the estimator, and they determine the temperature coefficient of the liquid parameter through the relation $S^{\mathrm{ex}} = -x_{\mathrm{Cr}} x_{\mathrm{Ni}} \sum_k b_k (x_{\mathrm{Cr}} - x_{\mathrm{Ni}})^k$. The use of 108-atom cells is accepted when the vibrational entropy at the equiatomic composition differs by less than 5\% from the value obtained with 864 atoms, a criterion that is met in this case with a difference of 0.45\% (\cref{fig:entropy}d). The excess electronic entropy, on the other hand, is assumed to be zero, since in a metallic liquid this contribution arises only from the change in the density of states at the Fermi level with respect to the pure liquids, and it is expected to be small in comparison with the uncertainty of the vibrational term.

\subsection{Objective and fitting sequence}
\label{sec:methods:objective}
\label{sec:methods:sequence}

We combine the evidence, the one-shot energies, and the entropy of the liquid into a single objective function, and the construction and optimization of this function is the part of the assessment most often left undocumented. In this work, we refine the model with PhaseForge+~\cite{sariturk2025phaseforgeplus}, maximizing the likelihood of ESPEI~\cite{bocklund2019espei} using \emph{analytical gradients of the CALPHAD model fitting error} obtained via the Jansson derivative method~\cite{kunselman2024analytically,kunselman2025analytical}, with pycalphad as the engine for calculating equilibria. Here we note that gradient-based CALPHAD  model optimization is truly enabling, as it reduces the CALPHAD model fit to a multi-dimensional gradient-based optimization problem where all the model parameters can be optimized at once, a radically different approach compared to the standard CALPHAD ad-hoc practice of piecewise model parameter fitting. It is also important to point out that this gradient-based CALPHAD fit~\cite{kunselman2025analytical,kunselman2026hmc} is also orders of magnitude faster than the standard MCMC approach in ESPEI~\cite{bocklund2019espei}.

Since the name of a software environment does not guarantee its contents, the version of ESPEI actually imported during each run is recorded in the run cards. Before the fit, we also compare the analytical gradients term by term with those of ESPEI 0.9.0, at the reference parameters and at a few random points, and with finite differences. Each observation enters the likelihood with the weight implied by its own uncertainty (\cref{sec:methods:orchid}), without multipliers set by hand. Even observations with no reported uncertainty are assigned the uncertainty that the rules of ORCHID would give them, rather than the software default.

Two kinds of observations require special treatment in the objective. Phase boundaries reported with only one end of the tie-line known are split into two: the known phase at the reported composition and, in addition, an equilibrium between this phase and the second phase at a vanishing phase fraction. This corresponds to the zero-phase-fraction condition used in ESPEI, which leaves the partner-phase composition to be determined by the model rather than by an estimate made by the assessor. Excluded phases, in turn, enter the objective through the hinge term of PhaseForge+, which penalizes an excluded phase whenever its Gibbs energy approaches the plane defined by the stable phases at the observed state within a margin $\tau$, set to \qty{100}{\J\per\mole} of atoms by default. Since the most certain tiers correspond to small energy scales, we smooth the onset of this penalty to prevent the optimizer from stalling. As noted earlier (\cref{sec:methods:orchid}), the energy scale of each exclusion is the larger of the certainty tier and the state uncertainty, converted to energy. We obtain the driving-force scale of each zero-phase-fraction record similarly, using the record's uncertainty and its tier as a lower bound.

The fit itself follows a sequence that starts at the liquid and moves down in temperature, in the same order in which experienced CALPHD assessors tend to approach a given system. This order reflects the fact that fewer phases compete at high temperatures, where the liquidus provides a well-defined set of constraints and where one-shot energetics, together with liquid entropy, already restrict the range of acceptable model parameters before the more crowded solid-state region is considered. In the first step, the liquid and the solid phases that melt are therefore fitted to the liquidus, the solidus and the melting invariants, as well as to the excess entropy of the liquid. In the second step, the calorimetric and activity data are added with their own uncertainties; in the last step, the phase-boundary data, the dense assemblages, and the excluded phases are added. Throughout the sequence, a held phase is never fixed. Instead, it remains free and is anchored to its one-shot value through an observation carrying the uncertainty of that value, since fixing a compound at its one-shot energy can leave it stable in regions where it has not been observed, as was the case for CrNi$_2$ in an early version of this assessment. To detect such problems early, the phase diagram is calculated and inspected (by AI agents) after every step, and any step that changes the phase diagram's topology is reported rather than continued.

Since CALPHAD model optimization objectives (i,e., the likelihood) usually have several local optima, each step of the sequence is fitted from multiple starting points, and each start must be polished to convergence before it can be compared to the others. A start is considered to have converged when the norm of the gradient, in the scaling used by the optimizer, falls below a threshold that is stated before the fit ($10^{-3}$ by default), or when it reaches a point at which none of the one-sided directional derivatives can improve the objective---the need for this second condition arises from the fact that the CALPHAD model fitting objective is only piecewise smooth: wherever an observation lies on a phase boundary of the model, the stable assemblage changes at that point, and it is then not possible for the norm of the gradient to become small, even at a true optimum. This same feature also tends to make the fitness landscape quite rugged. For this reason, we follow multistart with basin hopping~\cite{wales1997basinhopping} and dual annealing~\cite{xiang1997dualannealing}, as implemented in SciPy~\cite{virtanen2020scipy}, both starting from the best local result, with a fixed evaluation budget and a random seed set in advance. We record the complete trajectory of each start, including the objective, its individual terms, and the full parameter vector at every evaluation, along with the results; starts without this history are labeled as such, without attempting to reconstruct the missing trajectory by interpolation.

Among the converged starts, the fit carried forward to the next step is chosen by ranking rather than by a threshold on the objective function. Candidate energies are first filtered for plausibility, requiring compound energies to lie within about \qty{10}{\kilo\joule\per\mole} (atom) of the one-shot energies, thermochemical residuals to be on the order of their uncertainty, and pure-element limits to match those of SGTE. In this way, a \emph{fit that achieves the correct topology at the expense of implausible energies cannot be selected}. The plausible candidates are then ranked according to the topology of their phase diagrams, that is, the type of each reaction, the extent of each field, and the absence of spurious phases, with a spurious phase at practical temperatures counted as more serious than one far above the liquidus. Only after this are the remaining candidates ranked according to their quantitative agreement with the data, first in terms of the invariant temperatures and then in terms of the thermochemical residuals of each dataset. Ties between nearly equivalent candidates are resolved by the objective value and never by quantities read from a calculated grid.

\subsection{Physical checks and uncertainty}
\label{sec:methods:checks}
\label{sec:methods:uq}

The ranking described above can only choose among the models produced by the fit. A model with a higher likelihood, however, is not necessarily more physically sound. The general protocol therefore includes a set of checks that are independent of the objective. The \emph{first} of these checks is a gate on the pure elements, for which the phase sequence and melting points must coincide with those of the SGTE descriptions, with ordered end-members equal to their disordered parent. The \emph{second} check concerns the liquid, which must remain convex in composition wherever no liquid miscibility gap has been reported or suspected. The \emph{third} check concerns the possible appearance of a miscibility gap in the solid solution phases. Such a split is not penalized by the objective, since a single-phase observation is satisfied equally well by the two coexisting composition sets, and it can therefore appear in a fit without any cost. For this reason, in the absence of any report of a miscibility gap, and when the mixing energies do not suggest one, a convexity constraint is applied to the phase at its observed single-phase states. The constraint is imposed only above a temperature $T_{\mathrm{allow}}$, the highest temperature at which a metastable gap can be accepted, leaving a concavity that is real but metastable at lower temperatures to the data.

The temperature $T_{\mathrm{allow}}$ is set before the fit, based on two estimates of the critical temperature of a possible miscibility gap. The first estimate corresponds to the incoherent (stress-free) gap. In this case, the regular-solution limit $T_c \approx \Omega/2R$ (with $\Omega$ from the energy of mixing at the equiatomic composition) provides an upper bound, which is then corrected for short-range order and excess vibrational entropy. The second estimate corresponds to the coherent gap, and it is obtained by adding the elastic term of Cahn~\cite{cahn1961spinodal} to the curvature of the Gibbs energy, with elastic constants and lattice parameters calculated with the potential. We report both estimates, along with their uncertainty, derived from the potential error in $\Omega$. The coherent estimate, however, describes the decomposition of a coherent solid solution, not the stable equilibrium represented by the CALPHAD description. A gap that disappears once such corrections have been applied, and that no source reports, supports the use of the convexity constraint. A gap that survives them, on the other hand, remains a real possibility, and for this reason it is left to the data.

Once a model has been chosen, its uncertainty is estimated at low cost from the optima already found during the multistart. For this purpose, the converged starts from the final step are grouped into distinct modes by their parameters, with two starts assigned to different modes when their parameters differ by more than the reproduction tolerances, even if their objectives are nearly equal. This criterion matters because, as described above, the objective is nearly flat in some directions, so two distinct minima can have almost the same posterior value. At each mode, we compute the Hessian of the negative log posterior via central differences of the analytical gradient, then approximate the posterior as a mixture of Laplace approximations~\cite{tierney1986laplace}, with weights that depend on the posterior value at each mode and the determinant of the corresponding Hessian. Parameter vectors drawn from this mixture with a fixed seed are classified on a \qty{5}{\K} grid over the assessment temperature range, according to their topology, invariant temperatures, miscibility gaps, and spurious fields. This classification provides the probability of the accepted topology and its alternatives, along with their binomial errors, and allows the model uncertainty to propagate to any feature of the phase diagram. Markov chain Monte Carlo sampling~\cite{honarmandi2019bayesian,kunselman2026hmc}, which is much more expensive, is not part of the protocol by default and is used only when approved and recorded as a decision.

\subsection{Packaging and reproduction}
\label{sec:methods:package}

The fit, the checks and the estimate of uncertainty described so far all lead to a thermodynamic database, but under the proposed ORCHID-SAGE standard, an assessment is considered complete only when this database is delivered together with the documents that produced it. The database itself carries a header that states how, with which potential, and from which parent files it was generated, while its body remains byte-identical to the file produced by the fit.

Together with the database, the package contains the specific protocol, the run cards, the evidence bundle, and the assessment dashboard. These documents are assembled into a Workflow Run RO-Crate~\cite{soilandreyes2022rocrate,leo2024wrroc}, in which the specific protocol is the workflow that was actually run and the general protocol is the workflow on which it is based. The run cards, in turn, become actions with their instruments, inputs, outputs, and times, which can be mapped onto the vocabulary of PROV-O~\cite{lebo2013provo} in order to follow the provenance of any result. Inputs that are not part of the crate, such as restricted sources, appear in it only through their hashes and access conditions.

Of these documents, the dashboard is the one intended to make the record accessible to a person. It consists of a single standalone page that is rendered from the evaluation results document, and that can be read without access to a network. For each observed state, the dashboard shows whether the calculated phases correspond to those that were expected, and it also shows the trajectory of the starts in each step of the fit, the evolution of the fit in model space, the residuals of the datasets, the physical checks, the uncertainty, and the decisions taken along the way (\cref{fig:dashboard-objective,fig:dashboard-optimizer,fig:dashboard-parameters,fig:dashboard-projection}). To allow anyone to recompute the projection into model space from the recorded parameter vectors, this projection is deterministic. The parameters are first divided by stated scales and centered on the mean of the retained vectors, then projected onto the two leading eigenvectors of their covariance, with the sign of each eigenvector fixed by its largest component (\cref{fig:dashboard-projection}). Quantities not yet computed are shown as pending, with the reason, and are never filled in with estimates.

Although the dashboard is meant to be read by a person, reproducing a database requires a single command (\texttt{orchid\_reproduce.py}), which needs only the specific protocol and an empty directory. In the first stage, this command copies the code and the assessment evidence into the directory, checks every file against its recorded hash, and verifies that the interpreter matches the recorded environment. In the second stage, it executes the reproduction plan of the specific protocol and compares the new results with the original ones. As discussed in \cref{sec:methods:levels}, the verdict of this comparison is based on the predictions of the two databases, together with byte-for-byte checks of the identity of the data, while differences in the fitted parameters, in the environment, or in the labels of the starts are reported as notes. Steps that cannot be reproduced bitwise, such as relaxations and molecular dynamics simulations performed with GPUs or refits performed with a different processor architecture, are precisely those for which tolerances must be declared. Finally, an independent rerun of the complete specific protocol on a different machine is one of the gates a database must pass before release, and we report this rerun for Cr--Ni in \cref{sec:reproduction}.

\subsection{Machine execution through the Model Context Protocol}
\label{sec:methods:mcp}

The documents described so far can be read by a machine, but a record that a machine can read is not yet, necessarily, a record that a machine can carry out. In order to close this gap, the protocols, run cards, evidence bundles, validator, reproduction tools, and dashboards of ORCHID-SAGE are exposed through a server based on the Model Context Protocol~\cite{mcp}, alongside the existing ORCHID server for the data, so that an AI agent can use them as tools, resources, and prompts. In reproduction mode, an agent without prior context can list available protocols, read the steps of a specific protocol with exact commands, inputs, outputs, and tolerances, and verify that the inputs match the recorded hashes. The agent can then obtain a reproduction plan, run it one step at a time, and compare the outputs with those of the original assessment. Since some steps, such as the fits, take hours, they start in the background, and the agent follows their progress through a separate status call so no step depends on a single call remaining open. For a judgment step, on the other hand, the server returns the decision recorded in the specific protocol, together with its source, so that the agent applies the decision that was actually made instead of making one of its own.

In addition to reproducing existing assessments, the same server can start a new assessment for a system that has not yet been assessed. In this case, the server instantiates a draft of the specific protocol from the general protocol in a fresh directory and opens a request for each decision the general protocol leaves to a person, such as the choice of phases, the temperature range, and the source of the unary descriptions. It then gathers the evidence from the local ORCHID data, with certainty tiers and excluded phases at the observed states, and works through the steps of the general protocol in order. It runs computable steps, produces a job specification for steps that require a GPU, and pauses judgment steps until a person answers the corresponding request with a name and a reason the server cannot supply on its own. Because every step writes a run card, an assessment started this way is reproducible from its first step. However, in its present form, this mode has two limitations: fitting a new system still requires a person to adapt the reference code, which is then recorded as a change of tool, and there is not yet a dashboard for an assessment started this way.

In both modes, people remain free to change a protocol, provided that they document the change within the schema. A change to a specific protocol is recorded as a departure from the step of the general protocol it modifies, along with a decision stating who made the change, when, why, and on what evidence. Such a change produces a new version of the specific protocol, which must be validated before it is accepted; the steps affected by the change, as well as those that depend on them, are then marked as stale and rerun, while their earlier run cards are kept. A change to the general protocol, in turn, produces a new version of this protocol and never an edit of a released one, and an assessment moves to the new version only through a recorded decision. In this work, for example, correcting an error in the evidence bundle construction and a race condition between parallel workers led to new versions of both the evidence and the specific Cr--Ni protocol, while all earlier versions remained unchanged.

Since an agent acts on the record through these tools, the server has also been designed to prevent the agent from damaging the record it is reading. All runs take place in fresh directories that the server creates and marks, and these directories cannot be located inside the repository, the active campaigns, or the data copies. Each command is checked before it is executed, and a command that writes outside its directory, deletes files, or connects to another host is refused. In addition, the system runs a single step per call, after explicit confirmation that must be repeated for long steps such as fits. The server does not run steps that require a GPU, and the person must enable remote machines. Finally, none of the server tools can delete a file or access the network, and restricted sources are hashed to verify their presence but are never read out or copied.

\subsection{AI assistance and human decisions}
\label{sec:methods:ai}

The server safeguards are a particular case of a more general division of work between AI agents and (expert) humans, intended to leave the agents with steps that can be checked against the record and to reserve to people the decisions that cannot. In the assessment of Cr--Ni, AI agents (Claude Opus 5.5, Anthropic) working under the general protocol carried out most of the steps, from the search of the literature and the extraction and verification of the evidence to the one-shot calculations, the optimization of the model and the assembly of the specific protocol, the run cards and the dashboard. In contrast, the present author made the decisions that the protocol reserves to expert humans, including selecting phases, accepting new records, and adopting a new rule whenever a result revealed that one was missing. The protocol's decision log documents each of these decisions, including the date, the person responsible, the justification, and the source on which the decision is based whenever one exists. Decisions agents make on their own appear in the same log, where they are identified as such. Records extracted for this work from a single reading, finally, remain flagged until a second reader has verified them against the original source.

In this sense, the decision log also documents how the evidence has shaped the method. Several rules in the general protocol were introduced in response to specific results from this assessment and earlier ones. Examples include the requirement to specify excluded phases at all observed states, the rule that held parameters are anchored rather than fixed, the treatment of transforming structures through the midpoint of their relaxation interval and the lower bound on the energy scale of each exclusion. Each of these rules is now part of the standard, and the result that motivated it is recorded together with it. Thus, a protocol under ORCHID-SAGE constitutes a version of the method, one that evolves as new evidence requires and in which the assessor's judgment remains free, provided that the exercise of this judgment is documented.
\section{Results and Discussion}
\label{sec:results-and-discussion}

The validity of a reproducibility standard can only be established by regenerating an assessment from the record it prescribes, using no information beyond that contained in the record itself. The principal test of ORCHID-SAGE therefore consists of the independent regeneration of the Cr--Ni thermodynamic database from its specific protocol, and this test is presented before the assessment, since it constitutes the main evidence supporting the central claim of this work. Two independent regenerations were performed, one through scripted execution of the reproduction plan and one by an AI agent with access to the ORCHID-SAGE server tools and to read-only file tools, and in both cases the results were compared with those of the original assessment using the tolerances declared in the protocol (\cref{fig:verification}b).

\subsection{Reproduction of the Cr--Ni assessment}
\label{sec:reproduction}

\begin{figure}[tbp]
    \centering
    \includegraphics[width=83mm]{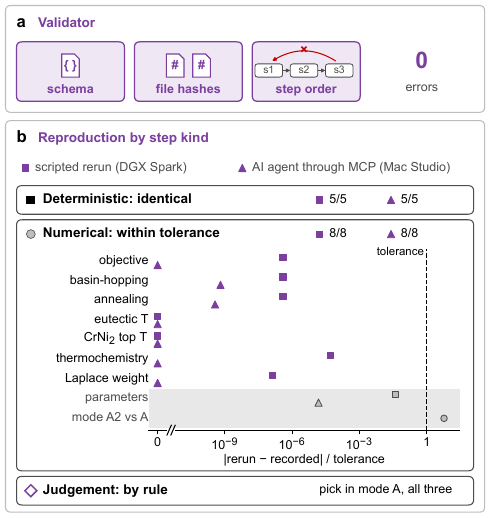}
    \caption{Verification of the specific protocol and of its reruns. (a)~The validator checks that every document conforms to the schema, that every referenced file exists with its recorded hash and that no step uses an input before it is produced. (b)~Two independent reruns, one scripted on a DGX Spark and one by an AI agent through the ORCHID-SAGE server on the Mac Studio, compared with the original by step kind: deterministic outputs are identical, numerical outputs are shown relative to their declared tolerances, and the judgment step picks a start in the same mode in all three runs. The parameters are compared only as a note (grey).}
    \label{fig:verification}
\end{figure}

\begin{figure*}[tbp]
    \centering
    \includegraphics[width=171mm]{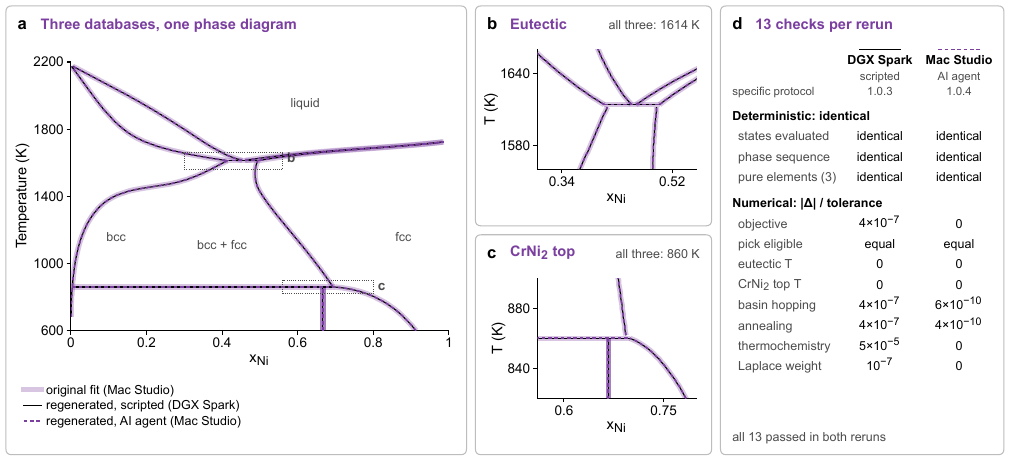}
    \caption{Regeneration of the Cr--Ni database from its record. (a)~Phase diagrams calculated from the database of the original assessment (broad light purple line) and from the databases regenerated by the scripted rerun on a DGX Spark (black line) and by the AI agent on the Mac Studio (dashed purple line). (b)~The eutectic and (c)~the top of CrNi$_2$, enlarged, with the temperature given by all three databases. (d)~The checks of the specific protocol for each rerun: deterministic checks give identical results, and numerical checks are shown as the difference from the recorded value relative to its tolerance.}
    \label{fig:repro}
\end{figure*}

The first regeneration was carried out with a script on an NVIDIA DGX Spark (aarch64, Linux), a machine that differs from the Mac Studio (arm64, macOS) used in the assessment in its processor, its operating system and its numerical libraries. Only the records files, identified by their SHA-256 hashes, were available on this machine. Of the 13 checks declared in the protocol, the five deterministic ones (the phases calculated at the 695 evaluated states, the phase sequence on heating at 99 compositions and the stable phases of the pure elements) gave identical results (\cref{fig:repro}). The eight numerical checks agreed within their tolerances, with a difference of $4 \times 10^{-9}$ in the objective (i.e. log likelihood) of the selected solution and no difference in the eutectic and CrNi$_2$-top temperatures (\cref{fig:verification}b). The selection rule picked a different start on this machine (s1 instead of s4), although both starts belong to the same posterior mode and lie 0.039 tolerances apart in parameter space. For this reason, the parameters are compared only as a note, while the model's predictions are compared within tolerances.

The second regeneration was carried out by an AI agent (Claude Opus 5.5), which had access to the tools of the ORCHID-SAGE server and to read-only file tools but received no information about the assessment other than the identifier of its specific protocol. Using only these tools, the agent read the protocol, staged its inputs, executed each of its steps through the server and compared the results to the recorded ones. This regeneration required 46 turns and 2~h~38~min, at a cost of US\$1.64 in model usage. Since it took place on the Mac Studio, the eight starts of the final step reproduced the original ones exactly, and all 13 checks were passed, eleven of them without any difference and the other two (the best results of basin hopping and dual annealing) with differences of the order of $10^{-12}$ in the objective. These (preliminary) results show that an AI agent can regenerate a CALPHAD database from its record alone, without human intervention, and this is precisely the condition a living database, updated every time new data appear, must satisfy. The same refit was finally repeated from a clean copy of the public repository of ORCHID-SAGE, in an environment built only from public sources (ESPEI, pycalphad and PhaseForge+ at their recorded commits), and the selected solution, the eutectic and the top of CrNi$_2$ were again reproduced exactly, in about one hour on a single workstation.

\begin{figure}[tbp]
    \centering
    \includegraphics[width=83mm]{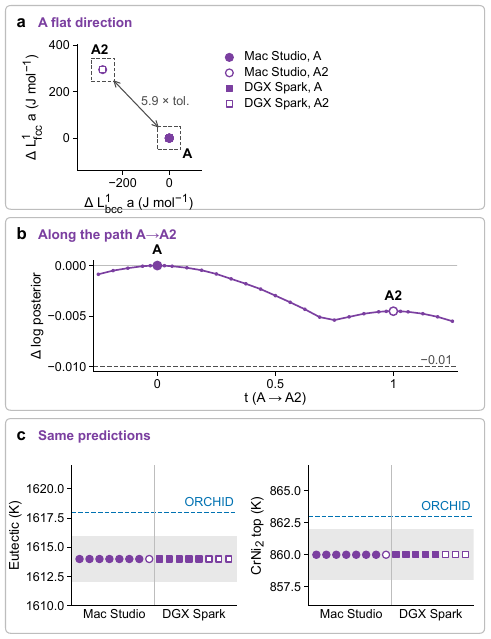}
    \caption{Two modes of the final fit with the same predictions. (a)~The two minima of the basin, A and A2, in the plane of the bcc and fcc first-order terms, 5.9 times the parameter tolerance apart (dashed boxes). (b)~Log posterior along the straight path from A to A2, which stays within the tolerance of 0.01 on the objective (dashed). (c)~Eutectic and CrNi$_2$-top temperatures at every start (filled, mode A; open, A2) on both machines, within \qty{\pm 2}{\K} of the recorded values (grey), with the ORCHID records (dashed). ORCHID v2026.10 and records extracted for this work; record identifiers are provided with the source data.}
    \label{fig:modes}
\end{figure}

In the final step of the fit for the Cr--Ni system, the objective was observed to have two minima of almost the same value (modes A and A2 in \cref{fig:modes}), which differ by only 0.0045 log units but by up to 5.9 times the parameter tolerance in the first-order Redlich--Kister terms of bcc and fcc. On the original machine, seven of the eight starts converged to mode A and one to A2, while on the second machine the split was five to three. Both modes, however, give identical phase diagrams, invariant temperatures and thermochemical properties. The data used in the fit therefore do not determine these two terms within the declared tolerance, and they are reported as undetermined in the record. In a conventional assessment, in which a single set of parameters is published, this would most likely have gone unnoticed.

\subsection{Cr--Ni: the worked case}
\label{sec:crni}

\begin{figure}[tbp]
    \centering
    \includegraphics[width=83mm]{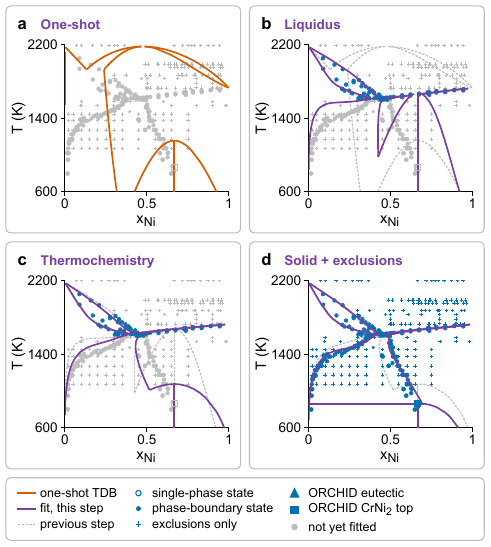}
    \caption{The Cr--Ni fit sequence, from the one-shot model to the final fit, calculated with pycalphad. (a)~The one-shot model of PhaseForge, without any fit. (b)~After the liquid step, fitted to the liquid phase-boundary data, the eutectic and the excess entropy of the liquid. (c)~After the thermochemical step, with CrNi$_2$ free. (d)~The final fit, with the solid-state data and the excluded phases at every observed state. Purple lines: the current step; dashed grey: the previous step. Markers: the 695 experimental states, in blue where the current step uses them and in grey where it does not yet. ORCHID v2026.10 and records extracted for this work; record identifiers are provided with the source data.}
    \label{fig:sequence}
\end{figure}

As mentioned above, the assessment starts with the one-shot model of PhaseForge~\cite{zhu2025accelerating,zhu2025machinelearningpotentialsalloys}, which combines potential-calculated energetics with SGTE descriptions of the pure elements. Although this model contains all phases of the system, its phase diagram is displaced at high temperatures (\cref{fig:sequence}a), and this error can be traced back to a single end-member. Fcc Cr is mechanically unstable, and its energy relative to bcc Cr is \qty{7.3}{\kilo\joule\per\mole} in the SGTE description but about \qty{38}{\kilo\joule\per\mole} according to the potential. Since the fcc mixing energies refer to fcc Cr in the potential, adding them to the SGTE end-member overstabilizes the Cr-rich fcc solution. As a consequence, the fcc field of the one-shot diagram extends to $x_{\mathrm{Ni}} \approx 0.26$ at \qty{1600}{\K}, bcc dissolves almost no Ni, and the eutectic between the liquid, bcc and fcc is located near \qty{1935}{\K}, instead of the \qty{1618}{\K} reported in ORCHID. Similarly, CrNi$_2$ remains stable up to \qty{1156}{\K}, well above the \qty{863}{\K} of the ORCHID record. The one-shot PhaseForge model therefore gives the correct set of phases and a reasonable first estimate of their energetics, although not a usable phase diagram; this is the starting point for refinement.

As we mention above, the refinement follows the order that is standard in CALPHAD practice~\cite{lukas2007}, in which the liquid and the phases that melt are fitted first, and the solid state is fitted last. The reason for this order is that fewer phases compete at high temperatures, where the liquidus and the solidus are usually the most abundant and reliable data. The parameters of the liquid can thus be determined before the more crowded solid-state region is considered, and they then serve as a stable reference for the rest of the fit. In the first step, we therefore fit the liquid and solid phases that melt to the liquidus, the solidus, and the eutectic, using the liquid excess entropy and the one-shot energies as additional constraints. This step brings the lowest temperature of the liquid down to \qty{1616}{\K} (\cref{fig:sequence}b), although it also leaves CrNi$_2$ stable up to \qty{1664}{\K}, since changes in the solution phases displace the field of a compound whose energy has not yet been adjusted.

In the second step, we add the thermochemical data to the fit, with CrNi$_2$ free but anchored to its one-shot energy via an observation with its uncertainty. The effect of this step is to lower the top of the compound to \qty{1074}{\K}, while the lowest temperature of the liquid remains at \qty{1626}{\K} (\cref{fig:sequence}c). In the last step, we add the phase-boundary data, along with the excluded phases at all observed states, and adjust the model's solid-state boundaries to the evidence. The resulting model places the eutectic at \qty{1614}{\K}, compared with \qty{1618}{\K} in the ORCHID record (with an effective uncertainty of \qty{11}{\K}), and the top of CrNi$_2$ at \qty{860}{\K}, compared with \qty{863}{\K} (\qty{16}{\K}) in a record that has so far been read by a single reader (\cref{fig:sequence}d). At each step, the model changed mainly in the region the newly added data could constrain: first in the liquid, then in the compound and the thermochemistry, and finally in the boundaries of the solid state, which is precisely the purpose of a sequence like this.

Two invariant temperatures are not enough to judge a phase diagram, and for this reason the final model was also compared with each of the 695 experimental states in the evidence bundle. For 599 of these states, the phases calculated at the reported temperature and composition include those observed; for 424, the calculated and observed assemblages are identical, while an excluded phase appears as stable in only 6 states. However, the extent to which a state is reproduced depends on how completely it was characterized in the original work. Complete tie-lines, for which both phases and both compositions are known, are reproduced in 40 of 41 cases, whereas this is the case for only 306 of the 489 single-phase states and 78 of the 164 partial tie-lines. In part, this difference is a consequence of the comparison itself. Since each state is evaluated at its reported coordinates and not within its uncertainty, a point close to a calculated boundary can easily fall on the wrong side of it, and these counts should be read as a lower bound of the agreement. For the thermochemical data, in contrast, the agreement is of the order of the reported uncertainties, with root-mean-square standardized residuals of 1.22 for the 113 activities and of 1.05 for the 130 enthalpies.

However, not all mismatches result from the comparison, and the largest ones in fact point to a real limitation of the model on the Cr-rich side of the diagram. According to the final model, sixteen single-phase bcc states reported in three independent studies~\cite{svechnikov1962characteristics,dench1963adiabatic,collins1988electron} lie inside the bcc + fcc or bcc + liquid fields. This means that the model places the boundary of the bcc solid solution at higher Cr contents than those observed (up to 0.088 in mole fraction at \qty{1403}{\K}, and 0.045 to 0.07 between \qty{1650}{\K} and \qty{1822}{\K}), that is, at four to eight times the uncertainty of the states. The same discrepancy appears in the residuals of these datasets, which are the largest of the fit. On the fcc side, on the other hand, it is the evidence itself that is inconsistent. Three fcc states reported by Jette et al.~\cite{jette1934xray} at $x_{\mathrm{Cr}} \approx 0.55$ and about \qty{1380}{\K} lie well beyond the fcc boundary of the model, while the fcc states of two other studies~\cite{collins1988electron,hao2021}
 at $x_{\mathrm{Cr}} = 0.40$--0.49 agree with it. We do not attempt to decide between these sources. However, both discrepancies are recorded in the assessment and point to the region of the diagram in which new measurements would be most useful.

\begin{figure*}[tb]
    \centering
    \includegraphics[width=171mm]{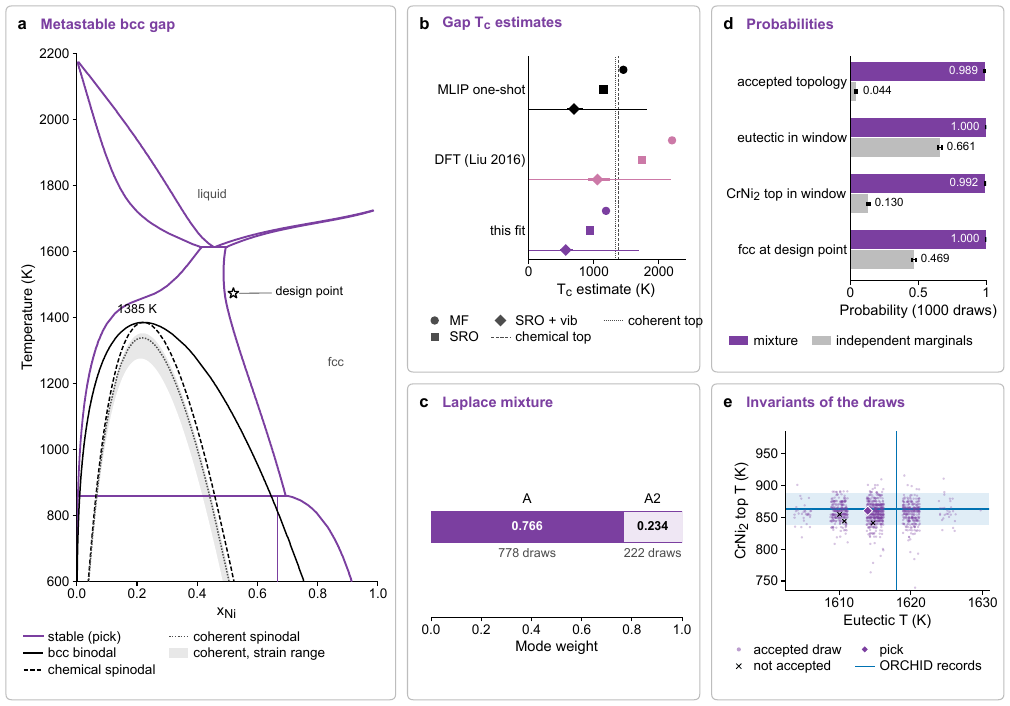}
    \caption{Physics and uncertainty of the final fit. (a)~Stable diagram (purple) with the metastable miscibility gap of BCC\_A2: binodal, chemical spinodal and coherent spinodal (grey band, range of coherency strain energies), all inside the stable two-phase fields; star, the design point at $x_{\mathrm{Ni}} = 0.52$ and \qty{1473}{\K}. (b)~Estimates of the critical temperature of the bcc gap as a regular solution, with short-range order and with the excess vibrational entropy, for the one-shot model, the DFT values of Liu et al.~\cite{liu2016alcocrni} and the fit. (c)~Weights of the two modes in the Laplace mixture. (d)~Probabilities of the accepted topology, the eutectic and CrNi$_2$ ranges and single-phase fcc at the design point, from the mixture (purple) and from independent marginals (grey). (e)~Eutectic and CrNi$_2$-top temperatures of the 1000 draws, with the ORCHID records (blue) and the final fit (diamond). ORCHID v2026.10 and records extracted for this work; record identifiers are provided with the source data.}
    \label{fig:physics}
\end{figure*}

The final model passes the physical checks described in \cref{sec:methods:checks}. The pure elements reproduce the SGTE descriptions, with bcc Cr and fcc Ni stable below their melting points, the liquid is convex in composition at all temperatures between \qty{600}{\K} and \qty{2200}{\K}, and no stable miscibility gap appears in either of the solid solutions. These checks also reveal a feature of the model that is not visible in the phase diagram: a metastable miscibility gap in the bcc solution. The chemical spinodal of this gap reaches \qty{1385}{\K} at $x_{\mathrm{Ni}} = 0.22$, and the elastic term of Cahn lowers the coherent spinodal to \qty{1338}{\K} (\cref{fig:physics}). Since the gap lies inside the bcc + fcc field, where fcc is the stable phase, it cannot be observed in equilibrium. The question of whether such a gap is reasonable cannot be settled with the energetics of the potential either. When short-range order and the excess vibrational entropy are taken into account, the incoherent estimate of the critical temperature obtained from the energetics of the potential is about \qty{700}{\K}, but the uncertainty of this estimate, which follows from the error of the potential, extends from zero to about \qty{1800}{\K}. Following \cref{sec:methods:checks}, we left the gap to the data, so its position in the final model is determined by the evidence rather than by one-shot energetics.

In order to estimate the uncertainty of the final model, the two modes of the flat basin were combined into a Laplace mixture, in which they enter with weights of 0.77 and 0.23, and 1000 parameter vectors drawn from this mixture were classified according to the procedure of \cref{sec:methods:uq}. It is important to note that 98.9\% of these draws give the accepted topology, while the eutectic remains within \qty{1610}{\K} to \qty{1620}{\K} and the top of CrNi$_2$ within \qty{830}{\K} to \qty{885}{\K} with a probability of 90\% (\cref{fig:physics}). This means the parameters left undetermined by the evidence have little effect on the model's predictions, consistent with the agreement between the two modes discussed in \cref{sec:reproduction}. This conclusion, however, depends on the correlations between the parameters. When the same marginal distributions are sampled independently, only 4.4\% of the draws give the accepted topology, and the top of CrNi$_2$ is spread over more than \qty{1000}{\K}. The Hessian-described correlations are thus an essential part of the uncertainty in a CALPHAD model, to the extent that neglecting them would grossly overestimate the uncertainty of the phase diagram.

\subsection{Discussion}
\label{sec:discussion}

The level of reproducibility demonstrated here has a cost, and it is therefore important to consider what is gained in exchange for it. Most of this cost is associated with the preparation of the record, which involves writing the general protocol, assembling the evidence together with its uncertainties and excluded phases, and documenting each run and each decision as the assessment proceeds. In this work, AI agents carried out much of this effort under the protocol, and the assessor's additional work was limited to the decisions the protocol reserves for people. In contrast, the cost of using the record once it exists is small. The complete regeneration of the Cr--Ni assessment by an AI agent through the server required 2~h~38~min on a single workstation, at a cost of US\$1.64 in model usage. In exchange for this cost, it becomes possible to verify any number in the database against the evidence and the decisions from which it originated, to detect errors in the record itself (as happened twice in the course of this work) and to extend the assessment as new data become available, without the need to reconstruct it from the beginning.

Some parts of an assessment, however, resist exact reproduction, and it is for these parts that the distinction between the kinds of steps of a protocol becomes important. Relaxations and molecular dynamics simulations performed with machine-learned potentials on GPUs are not bitwise deterministic, and an optimizer's trajectory depends on processor architecture, as shown by the different proportions at which the Cr--Ni fit reached its two modes on the two machines. For such steps, the standard requires agreement within tolerances that are declared in advance and based on the spread observed between repeated runs. A check of the liquid simulations on two DGX Sparks shows why the spread has to be measured rather than assumed. Runs started with the same seed did not reproduce their volumes even on the same GPU, and when the volume of the cell was averaged over only \qty{5}{\ps}, the enthalpies of the individual states differed between the two machines by up to \qty{3.5}{\kilo\joule\per\mole} of atoms, which is considerably more than their statistical errors would suggest. With an averaging window of \qty{50}{\ps}, in contrast, every state agreed to within twice its standard error. The enthalpy of mixing of the equiatomic liquid agreed between the machines within this tolerance for all three windows, with values between $-2.3$ and \qty{-4.4}{\kilo\joule\per\mole} of atoms compared with \qty{-3.2}{\kilo\joule\per\mole} of atoms in the assessment, and this spread remains well below the uncertainty of \qty{9.7}{\kilo\joule\per\mole} of atoms that is assigned to these values in the fit. In the case of the fit, these tolerances apply to the predictions of the model rather than to its parameters, because parameters in the nearly flat directions of the objective can change considerably with no effect on the phase diagram. An important consequence of this choice is that two assessments are regarded as equivalent when they predict the same phase diagram and thermochemistry, even when their parameters differ. The parameters that the evidence leaves undetermined, in turn, are reported as such, instead of being presented as quantities known to the precision with which they are printed.

A related question is why the protocols of ORCHID-SAGE were not written in an existing workflow language, such as CWL~\cite{crusoe2022cwl}. The reason is that a CALPHAD assessment is only partly a computation. A large fraction of its steps consists of judgments (the choice of phases and models, the acceptance of records, the selection of a fit among several candidates), while other steps require GPUs or campaign-specific code that a workflow engine would not run in a portable way. For this reason, the protocols record the assessment plan and the decisions taken along it, and refer to the exact command and code that were used for each executable step. This design does not exclude a workflow language, however. The deterministic and numerical steps of a specific protocol could be exported to CWL and executed by any compliant engine, with the judgment steps supplied from the decision log, and such an export would make the standard compatible with the infrastructure for workflows that already exists in other fields.

A complete record also brings to light questions about the model that would otherwise go unnoticed, and the metastable bcc gap described in \cref{sec:crni} is a case in point. In the final model, the chemical spinodal of this gap reaches \qty{1385}{\K}. The incoherent estimate of the critical temperature obtained from the fitted interaction parameter of bcc, in contrast to the one obtained from the energetics of the potential, and corrected for short-range order and for the excess vibrational entropy, is close to \qty{570}{\K}. The experimental data therefore imply a considerably larger gap in the model than the corrected energetics suggest. The information contained in the present record is not sufficient to explain this difference, and we do not attempt to explain it here. The point to be made is rather that such a difference can only be identified when the energetics, the corrections and the fitted parameters are part of a single record, and when the rule according to which the gap was left to the data has been stated before the fit was carried out.

The Cr--Ni assessment is, of course, only a first test of the standard. Binary systems with ordered phases, with compounds of wide homogeneity range or with magnetic contributions, as well as ternary systems in which several binary descriptions have to be combined, will place demands on the general protocol that the present case did not, and we expect the protocol to grow, version by version, as these systems are assessed. The evidence is not in its final form either. Some of the data used in this work were extracted by a single reader and are not yet part of a released version of ORCHID, so that they remain flagged in the assessment until a second reading confirms them. The reproduction reported here, finally, starts from the one-shot energetics, since the calculations with the machine-learned potential were only checked for a small sample of structures on a second GPU. A full regeneration of these calculations is possible with the same protocol, although it would take considerably more computing time than the rest of the assessment.

Beyond these limitations, the most important consequence of a standard of this kind is that \emph{it allows a thermodynamic database to take part in the loop between prediction and experiment} described in the \textbf{Introduction}. A database that can be regenerated from its record can also be updated when new evidence appears, by adding the new observations to the evidence bundle and running the specific protocol again, with every decision that is not affected by the new data kept as it was. In a self-driving laboratory~\cite{abolhasani2023selfdriving,szymanski2023alab}, the phase stability predicted by such a database, together with its propagated uncertainty, can guide the choice of the next experiments, and the observations made in these experiments can be recorded in ORCHID as fully defined thermodynamic states and used to refine the database.

Since an \textbf{AI agent can carry out the refinement} through the server, each version of this living database remains traceable to the evidence and the decisions from which it was obtained. The Cr--Ni assessment reported here is a first step in this direction, and the Cr-rich region of the bcc solid solution identified in \cref{sec:crni}, where the model and evidence disagree, is precisely the kind of place to which such a laboratory would direct its next experiments.
\section{Conclusions}
\label{sec:conclusion}

In this work, a standard for the reproducible assessment of CALPHAD databases, ORCHID-SAGE, has been introduced and tested on the Cr--Ni system. The main idea of the standard is that the record of an assessment has to include not only its data and code, but also the decisions of the assessor, and for this reason an assessment is recorded at three levels, namely the general protocol of the method, the specific protocol of the system and the run cards of the individual executions. The complete record is distributed together with the database as an RO-Crate. In the case of Cr--Ni, this record was sufficient to regenerate the database on a second machine, with the predictions of the model reproduced within the declared tolerances. Only the parameters along nearly flat directions of the objective differed between the reruns, and these parameters are reported in the record as undetermined by the evidence.

Beyond reproducibility, the standard is a first step towards CALPHAD assessments carried out by AI agents, since the expert knowledge that guides an assessment is encoded in protocols that a machine can follow. Since each protocol is written in a schema that a machine can read and each step is linked to the exact command that executes it, an agent with no prior knowledge of an assessment can read the protocol, run its steps and compare the results with the original ones. The rerun of the Cr--Ni assessment through the ORCHID-SAGE server provides direct evidence of this, since an agent with access only to the tools of the server and to read-only file tools regenerated the database without any other information and reproduced its predictions within the declared tolerances. The same server can also be used to start new assessments, in which the agent gathers the evidence from ORCHID and works through the general protocol step by step, although the fit steps for a new system are still adapted by a person, and the decisions that the protocol reserves for people are requested from them and recorded together with their reasons. ORCHID-SAGE thus provides a framework for AI-driven CALPHAD assessments that remain open to inspection, because the actions of the agents and the decisions of the people involved all become part of the record from which the database can be regenerated. Human guidance, however, remains essential in such assessments. Choosing phases and models, accepting new evidence, and adopting new rules require scientific judgment that agents do not replace, and the standard's role is to ensure this judgment is exercised explicitly and recorded, rather than left implicit in the final parameters.

The assessment itself, which combines the evidence of ORCHID with the one-shot model of PhaseForge and the gradient-based refinement of PhaseForge+, reproduces the measured eutectic and the top of CrNi$_2$ to within a few kelvin. It also points to the Cr-rich boundary of the bcc solid solution as the region of the diagram in which the model and the evidence still disagree, and in which new measurements would therefore be most useful. These results show that the reproducibility of a CALPHAD assessment is within reach at a modest cost, provided that the decisions of the assessor are recorded with the same care as the data. A database maintained in this way is no longer fixed at the time of its publication, and it can instead become a living database, refined as self-driving laboratories and other sources provide new evidence.
\section*{Author contributions}
\textbf{Raymundo Arróyave:} Conceptualization, Methodology, Software, Validation, Formal analysis, Investigation, Data curation, Writing -- original draft, Writing -- review \& editing, Visualization, Supervision, Funding acquisition.

\section*{Conflicts of interest}
There are no conflicts to declare.

\section*{Declaration of generative AI and AI-assisted technologies in the manuscript preparation process}
During the preparation of this work, the author used Claude Opus 5.5 (Anthropic) to draft the outline and text of the manuscript (under human supervision at the paragraph/sentence level) and to prepare the figures from the recorded results. Afterward, the author reviewed and edited every sentence of the manuscript and takes full responsibility for the content of the published article.
\section*{Data, code and reproducibility}

\paragraph{Data availability.}
The thermodynamic records used in this work are from ORCHID release v2026.10 (doi:10.5281/zenodo.23736065), released under CC BY 4.0 at \url{https://data.orchid-project.org}, together with records extracted for this work that are not yet part of a release. Each plotted or fitted value carries a record id (ORC-\ldots) and the id of the equilibrium state it observes (S-\ldots). These ids are the released ones for the records of v2026.10, and are otherwise derived with the identifier rules of ORCHID or, for the records extracted for this work, assigned in the provisional namespace \texttt{local:0000-0001-7548-8686/\allowbreak 93485c57-\allowbreak c1d0-\allowbreak 4264-\allowbreak bd60-\allowbreak 93a56a517d0a}, in which they become released ids by alias. The records extracted for this work are archived, with their sources, in the evidence bundle of the assessment, from which they can be cited. The figures and assessment evidence are described in the main text and are included with their source data in the paper repository. Each record retains its citation and source location in the assessment evidence bundle. Publications that cannot be redistributed enter the evidence only through derived values and are identified by their hashes.

\paragraph{Reproducibility.}
The CALPHAD database \texttt{step4c\_b\_hingefloor\_pick.tdb} was produced by the specific protocol \texttt{crni-paper-fit-step4c-b@1.0.4}, which instantiates \texttt{orchid-calphad-fit@2.1.2}. The protocol, run cards, evidence bundle, and assessment dashboard are archived as a Workflow Run RO-Crate (doi:10.5281/zenodo.24045965), and the protocol records the SHA-256 hashes of the TDB and all its inputs, along with the command that reproduces it. The dashboard is a single self-contained file, \texttt{dashboard/dashboard.html} in the RO-Crate, which can be opened in a web browser without a network connection. ORCHID-SAGE, which comprises the general and specific protocols, the schema, the validator, the reproduction tool and the MCP server, is released with this paper under the MIT license for code and CC BY 4.0 for protocols, schema and data, at \url{https://github.com/orchid-calphad/orchid-sage} (doi:10.5281/zenodo.24045967). The fits used PhaseForge+~\cite{sariturk2025phaseforgeplus} at a commit whose source tree is identical to that of commit 2da1d69 of its public repository (\url{https://github.com/dogusariturk/PhaseForgePlus/pull/163}), and the gradient-based branch of ESPEI at commit 9039ea2 (version 0.8.11.dev82), both of which are recorded in the run cards. The scripts and data behind every figure are available in the repository of this paper (\url{https://github.com/ArroyaveLab/orchid-sage-crni-paper}, doi:10.5281/zenodo.24045969).
\section*{Acknowledgments}
This manuscript is dedicated to the memory of Dr. Larry Kaufman (1931-2013), the Grandfather and \textbf{Sage} of CALPHAD and a mentor to this author and to many members of the CALPHAD community.

This work was supported by the U.S. National Science Foundation under Award OAC-2608825 (CSSI: Elements: PhaseForge---Building an Open, Reproducible Infrastructure for Machine-Learning-Assisted CALPHAD Modeling) and Award DMR-2534344 (ARM-MIP). Portions of this research were conducted with the advanced computing resources provided by Texas A\&M High Performance Research Computing and with the VISION supercomputer (NVIDIA DGX SuperPOD) of the Texas A\&M University System. The research was carried out interactively with Claude Opus 5.5 (Anthropic), which was also used to draft the outline and the text of the manuscript and to prepare the figures from the recorded results; the author reviewed and edited every sentence of the manuscript and takes full responsibility for its content. The AI agent that regenerated the Cr--Ni database is described in \cref{sec:reproduction}, and its complete log, with the model identifier and date, is part of the archived record.
\appendix
\renewcommand{\thefigure}{\thesection\arabic{figure}}
\makeatletter
\@addtoreset{figure}{section}
\makeatother
\section{Pair entropy of chemical mixing in the liquid}
\label{app:entropy}

\begin{figure}[tbp]
    \centering
    \includegraphics[width=83mm]{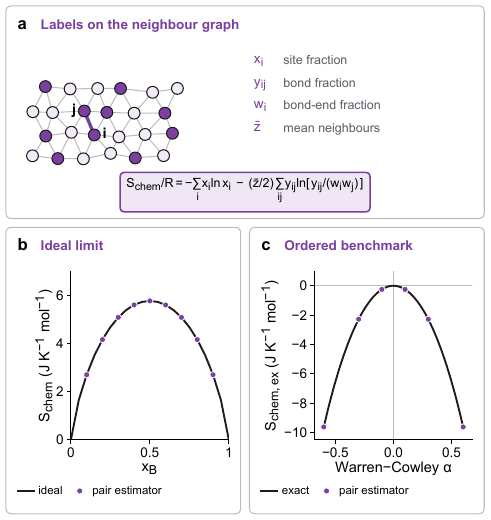}
    \caption{The pair estimator of the chemical-mixing entropy and its tests, in \unit{\J\per\K\per\mole} of atoms. (a)~Schematic neighbor graph with the site, bond and bond-end fractions that enter the pair (quasi-chemical) estimator. (b)~Random labels on a lattice with $z = 12$: the pair estimator (circles) recovers the ideal entropy of mixing (line) to machine precision. (c)~Bethe lattice with $z = 12$ at $x_B = 0.5$: excess chemical entropy against the Warren--Cowley parameter, exact (line) and from the pair estimator (circles). The derivation awaits independent review.}
    \label{fig:pairentropy}
\end{figure}

In the method of Hong and Liu~\cite{hong2025entropy}, the entropy of a liquid simulated by molecular dynamics is separated into a vibrational part, obtained from the velocity autocorrelation function, and a configurational part that accounts for the arrangement of the chemical species. The configurational part can be expressed as the entropy of the species labels on the neighbor graph of the simulation, in which two atoms are connected when their distance is smaller than the first minimum of the corresponding partial radial distribution function (\cref{fig:pairentropy}a). Whatever the estimator, this entropy has a known limit. When the labels are distributed at random over the sites of the graph, it must reduce to the ideal entropy of mixing, $-R\sum_i x_i \ln x_i$, independently of the structure of the graph. In this work, the chemical-mixing term of Hong and Liu~\cite{hong2025entropy} is replaced with the pair approximation described below, which satisfies this limit exactly (\cref{fig:pairentropy}b).

In the pair (quasi-chemical) approximation~\cite{guggenheim1944mixtures,kikuchi1951cooperative}, the entropy per site of a lattice with coordination $z$ is obtained from the entropy of the bonds minus the entropy of the sites that each bond counts twice, $S/(N k_{\mathrm{B}}) = (z/2)\,H(y) - (z-1)\,H(x)$, where $H$ denotes the Shannon entropy, $x_i$ the fraction of species $i$ and $y_{ij}$ the fraction of bonds with $i$ at one end and $j$ at the other. For the neighbor graph of a liquid, in which the coordination varies from atom to atom, we write the same expression in terms of the mean coordination $\bar z$ and of the bond-end fractions $w_i = \sum_j y_{ij}$, which reduce to $x_i$ on a lattice:
\begin{equation}
\frac{S_{\mathrm{chem}}}{R} = -\sum_i x_i \ln x_i - \frac{\bar z}{2}\sum_{ij} y_{ij}\ln\frac{y_{ij}}{w_i w_j}.
\end{equation}
The second term is $-(\bar z/2)$ times the mutual information $I$ of the species at the two ends of a bond, so that the excess chemical entropy is $S^{\mathrm{ex}}_{\mathrm{chem}} = -R\,(\bar z/2)\,I$.

This form has three properties that make it suitable as an estimator. Since $I = 0$ for random labels on any graph, the entropy reduces exactly to the ideal entropy of mixing in this limit, and since $I \ge 0$, any chemical order, whether ordering or clustering, lowers the entropy. In addition, the expression is exact for the quasi-chemical distribution on a Bethe lattice and for a one-dimensional chain, and it remains an approximation in other cases. The estimator was tested by permuting the species labels on the neighbor graphs of the Cr--Ni simulations, which gives excess chemical entropies of the order of $10^{-3}$~\unit{\J\per\K\per\mole} (\cref{fig:entropy}b), and on ordered benchmarks with Warren--Cowley parameters between $-0.6$ and $0.6$, for which it reproduces the exact excess entropy of the Bethe lattice (\cref{fig:pairentropy}c). The effect of the neighbor cutoff is included in the uncertainty of the estimate. Until the derivation has been reviewed independently, the excess entropies obtained with this estimator are treated as provisional. The estimator and its tests are available in the package \texttt{pair-entropy}~\cite{pairentropy}.

\section{Assessment trajectory and calculated phase states}
\label{app:assessment}

\begin{figure}[tbp]
    \centering
    \includegraphics[width=83mm]{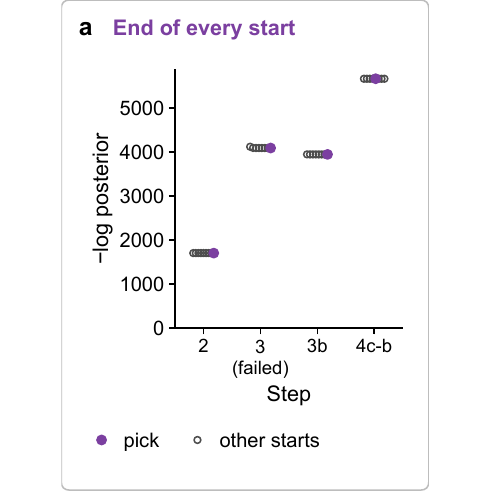}
    \caption{Negative log posterior at the end of each start in steps 2, 3,
    3b, and 4c-b; the selected start is purple. Objective values across steps
    are comparable only within each step because the evidence changes.}
    \label{fig:dashboard-objective}
\end{figure}

\begin{figure}[tbp]
    \centering
    \includegraphics[width=83mm]{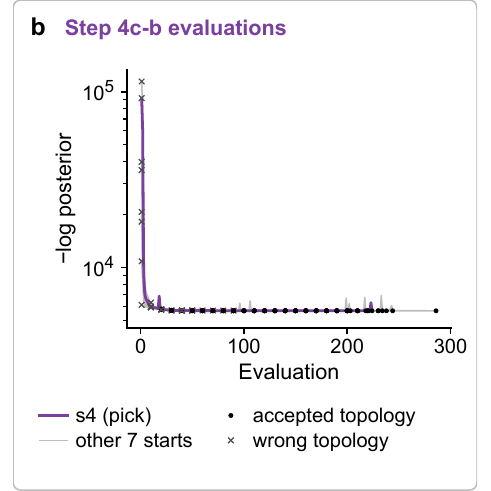}
    \caption{Evaluations of the eight step-4c-b starts; topology was
    classified every 10 evaluations.}
    \label{fig:dashboard-optimizer}
\end{figure}

\begin{figure}[tbp]
    \centering
    \includegraphics[width=83mm]{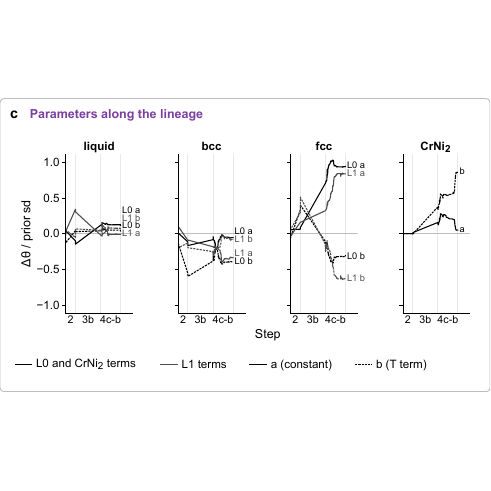}
    \caption{Changes in the 14 parameters along the selected lineage, in
    units of the prior standard deviation.}
    \label{fig:dashboard-parameters}
\end{figure}

\begin{figure}[tbp]
    \centering
    \includegraphics[width=83mm]{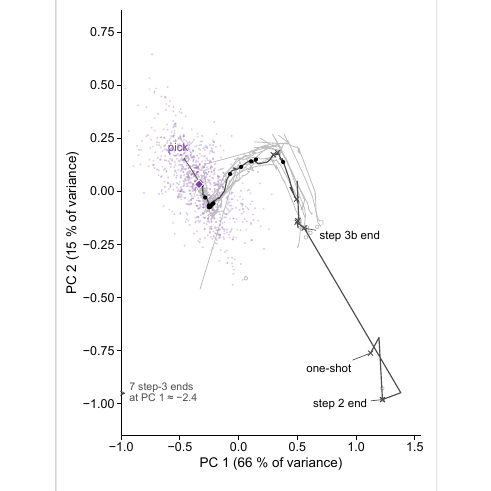}
    \caption{Projection of starts, lineage, and uncertainty samples onto the
    fixed PCA basis used by the dashboard. The purple points are 1,000
    Laplace-mixture draws and the purple diamonds are the mode centres. The
    dark line traces the selected lineage, black dots mark accepted topology,
    and crosses mark topology violations. Grey lines show the other step-4c-b
    starts, and open circles show the other starts. Seven
    failed step-3 endpoints lie outside the plotted range, at a mean PC 1 of
    approximately $-2.4$.}
    \label{fig:dashboard-projection}
\end{figure}

The assessment dashboard records how the candidate solutions change during
the fitting sequence and where the selected fit lies in parameter space. To
make its four views readable at column width, we show them separately in
\cref{fig:dashboard-objective,fig:dashboard-optimizer,fig:dashboard-parameters,fig:dashboard-projection}.
Panel (a) gives the negative log posterior for each start; because evidence is
added at each step, these values are comparable only within a step. Panel (b)
follows the eight starts of step 4c-b against the topology criterion. Panel
(c) shows how the 14 parameters change along the selected lineage, scaled by
their prior standard deviations, while panel (d) projects the starts, lineage,
and Laplace-mixture draws onto the dashboard's fixed PCA coordinates.

\begin{figure}[tbp]
    \centering
    \includegraphics[width=83mm]{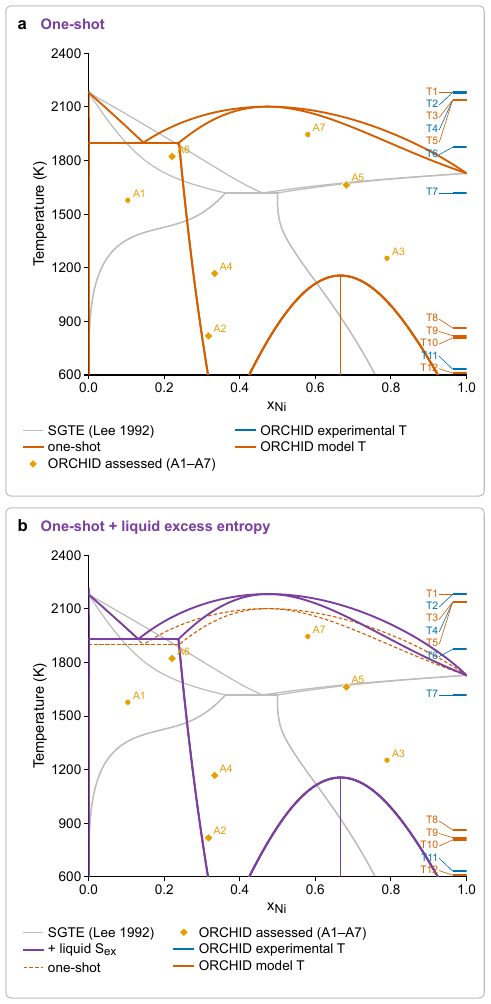}
    \caption{Phase diagram of the PhaseForge one-shot model of Cr--Ni,
    calculated with pycalphad, (a) without and (b) with the excess entropy of
    the liquid. Orange lines show the one-shot model and purple lines show the
    model with the liquid excess entropy; orange dashed lines in (b) reproduce
    panel (a). Light grey lines show the assessment of Lee
    \cite{lee1992}. Orange markers (A1--A7) denote assessed assemblages in
    ORCHID, with circles for single-phase fields and diamonds for two-phase
    fields. Marks at the right edge (T1--T12) denote transition and invariant
    temperatures without an assigned composition; blue marks are experimental
    and orange-red marks are model values. Records naming Al--Cr--Ni phases are
    excluded. The plotted observations are from ORCHID v2026.10 and the sources
    cited in the main text.}
    \label{fig:oneshot}
\end{figure}

The one-shot calculation provides the starting description for the refinement
discussed in the main text. In \cref{fig:oneshot}, panel (a) shows the
PhaseForge model before adding the excess entropy of the liquid, and panel (b)
shows the change after it is included. The Lee assessment \cite{lee1992} and
observations compiled in ORCHID provide a reference for both calculations.
The markers distinguish phase assemblages from transition temperatures, and
experimental transition temperatures from values reported by models. This
comparison shows what the liquid entropy changes before the remaining
differences are addressed by fitting.

\begin{figure}[tbp]
    \centering
    \includegraphics[width=83mm]{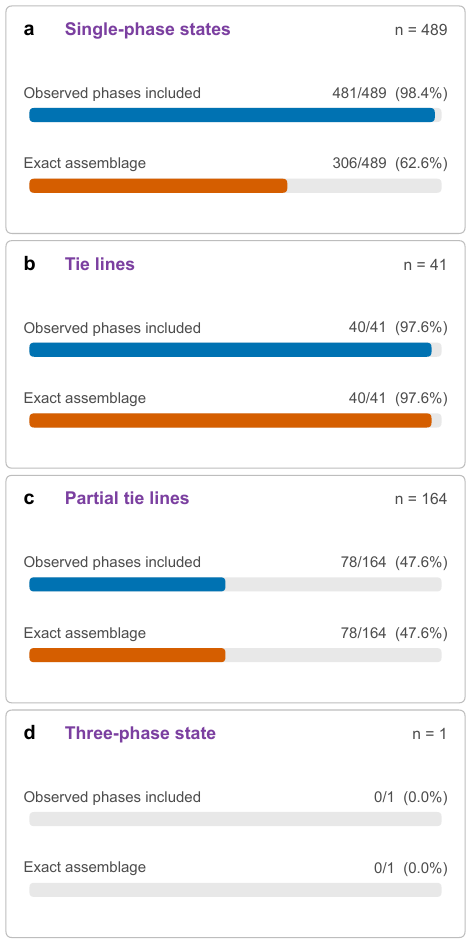}
    \caption{Comparison of calculated and observed phase assemblages by
    evidence type for the 695 states. Blue bars show the fraction for which
    all observed phases occur in the calculated assemblage; orange bars show
    exact assemblage agreement. Counts and percentages are given at right.}
    \label{fig:state-match-summary}
\end{figure}

We checked the final assessment at 695 observed states, comprising 329
phase-boundary records and 366 states with information about phases absent
at equilibrium. At each reported temperature and composition, we calculated
the stable assemblage at 101325 Pa and compared it with the observed phases.
A partial tie line reports one end of an equilibrium, leaving the composition
of the coexisting phase unspecified. As shown in
\cref{fig:state-match-summary}, the calculated assemblage contains all
observed phases at 599 states, while exact agreement occurs at 424. These
measures differ because inclusion allows the calculation to predict
additional phases. The comparison is made at the reported coordinates and
does not account for measurement uncertainty; the main text discusses how
this affects its interpretation.

\clearpage
\bibliography{main}

\begin{thebibliography}{50}
\providecommand{\natexlab}[1]{#1}
\providecommand{\url}[1]{\texttt{#1}}
\expandafter\ifx\csname urlstyle\endcsname\relax
  \providecommand{\doi}[1]{doi: #1}\else
  \providecommand{\doi}{doi: \begingroup \urlstyle{rm}\Url}\fi

\bibitem[Lukas et~al.(2007)Lukas, Fries, and Sundman]{lukas2007}
Hans~Leo Lukas, Suzana~G. Fries, and Bo~Sundman.
\newblock \emph{Computational Thermodynamics: The {Calphad} Method}.
\newblock Cambridge University Press, 2007.
\newblock \doi{10.1017/CBO9780511804137}.

\bibitem[Otis and Liu(2017)]{otis2017pycalphad}
Richard Otis and Zi-Kui Liu.
\newblock pycalphad: {CALPHAD}-based computational thermodynamics in {Python}.
\newblock \emph{Journal of Open Research Software}, 5\penalty0 (1):\penalty0 1, 2017.
\newblock \doi{10.5334/jors.140}.

\bibitem[Arr{\'o}yave(2022)]{arroyave2022phase}
Raymundo Arr{\'o}yave.
\newblock Phase stability through machine learning.
\newblock \emph{Journal of Phase Equilibria and Diffusion}, 43\penalty0 (6):\penalty0 606--628, 2022.
\newblock \doi{10.1007/s11669-022-01009-9}.

\bibitem[Bocklund et~al.(2019)Bocklund, Otis, Egorov, Obaied, Roslyakova, and Liu]{bocklund2019espei}
Brandon Bocklund, Richard Otis, Aleksei Egorov, Abdulmonem Obaied, Irina Roslyakova, and Zi-Kui Liu.
\newblock Espei for efficient thermodynamic database development, modification, and uncertainty quantification: application to cu--mg.
\newblock \emph{MRS Communications}, 9\penalty0 (2):\penalty0 618--627, 2019.

\bibitem[Arr{\'o}yave et~al.(2026)Arr{\'o}yave, Saal, and Shin]{arroyave2026orchid}
Raymundo Arr{\'o}yave, James~E. Saal, and Dongwon Shin.
\newblock {ORCHID}: an open, {AI}-ready resource of {CALPHAD} literature data, calculated phase diagrams, phase identities and {DFT} references, 2026.

\bibitem[M{\"o}lder et~al.(2021)M{\"o}lder, Jablonski, Letcher, Hall, Tomkins-Tinch, Sochat, Forster, Lee, Twardziok, Kanitz, Wilm, Holtgrewe, Rahmann, Nahnsen, and K{\"o}ster]{molder2021snakemake}
Felix M{\"o}lder, Kim~Philipp Jablonski, Brice Letcher, Michael~B. Hall, Christopher~H. Tomkins-Tinch, Vanessa Sochat, Jan Forster, Soohyun Lee, Sven~O. Twardziok, Alexander Kanitz, Andreas Wilm, Manuel Holtgrewe, Sven Rahmann, Sven Nahnsen, and Johannes K{\"o}ster.
\newblock Sustainable data analysis with {Snakemake}.
\newblock \emph{F1000Research}, 10:\penalty0 33, 2021.
\newblock \doi{10.12688/f1000research.29032.2}.

\bibitem[Di~Tommaso et~al.(2017)Di~Tommaso, Chatzou, Floden, Barja, Palumbo, and Notredame]{ditommaso2017nextflow}
Paolo Di~Tommaso, Maria Chatzou, Evan~W. Floden, Pablo~Prieto Barja, Emilio Palumbo, and Cedric Notredame.
\newblock {Nextflow} enables reproducible computational workflows.
\newblock \emph{Nature Biotechnology}, 35\penalty0 (4):\penalty0 316--319, 2017.
\newblock \doi{10.1038/nbt.3820}.

\bibitem[Crusoe et~al.(2022)Crusoe, Abeln, Iosup, Amstutz, Chilton, Tijani{\'c}, M{\'e}nager, Soiland-Reyes, Gavrilovi{\'c}, Goble, and {The CWL Community}]{crusoe2022cwl}
Michael~R. Crusoe, Sanne Abeln, Alexandru Iosup, Peter Amstutz, John Chilton, Neboj{\v{s}}a Tijani{\'c}, Herv{\'e} M{\'e}nager, Stian Soiland-Reyes, Bogdan Gavrilovi{\'c}, Carole Goble, and {The CWL Community}.
\newblock Methods included: Standardizing computational reuse and portability with the {Common Workflow Language}.
\newblock \emph{Communications of the ACM}, 65\penalty0 (6):\penalty0 54--63, 2022.
\newblock \doi{10.1145/3486897}.

\bibitem[Huber et~al.(2020)Huber, Zoupanos, Uhrin, Talirz, Kahle, H{\"a}uselmann, Gresch, M{\"u}ller, Yakutovich, Andersen, Ramirez, Adorf, Gargiulo, Kumbhar, Passaro, Johnston, Merkys, Cepellotti, Mounet, Marzari, Kozinsky, and Pizzi]{huber2020aiida}
Sebastiaan~P. Huber, Spyros Zoupanos, Martin Uhrin, Leopold Talirz, Leonid Kahle, Rico H{\"a}uselmann, Dominik Gresch, Tiziano M{\"u}ller, Aliaksandr~V. Yakutovich, Casper~W. Andersen, Francisco~F. Ramirez, Carl~S. Adorf, Fernando Gargiulo, Snehal Kumbhar, Elsa Passaro, Conrad Johnston, Andrius Merkys, Andrea Cepellotti, Nicolas Mounet, Nicola Marzari, Boris Kozinsky, and Giovanni Pizzi.
\newblock {AiiDA} 1.0, a scalable computational infrastructure for automated reproducible workflows and data provenance.
\newblock \emph{Scientific Data}, 7:\penalty0 300, 2020.
\newblock \doi{10.1038/s41597-020-00638-4}.

\bibitem[Draxl and Scheffler(2019)]{draxl2019nomad}
Claudia Draxl and Matthias Scheffler.
\newblock The {NOMAD} laboratory: from data sharing to artificial intelligence.
\newblock \emph{Journal of Physics: Materials}, 2\penalty0 (3):\penalty0 036001, 2019.
\newblock \doi{10.1088/2515-7639/ab13bb}.

\bibitem[Lebo et~al.(2013)Lebo, Sahoo, and McGuinness]{lebo2013provo}
Timothy Lebo, Satya Sahoo, and Deborah McGuinness.
\newblock {PROV-O}: The {PROV} ontology.
\newblock W3C Recommendation, 30 April 2013, 2013.
\newblock URL \url{https://www.w3.org/TR/2013/REC-prov-o-20130430/}.

\bibitem[Soiland-Reyes et~al.(2022)Soiland-Reyes, Sefton, Crosas, Castro, Coppens, Fern{\'a}ndez, Garijo, Gr{\"u}ning, La~Rosa, Leo, {\'O}~Carrag{\'a}in, Portier, Trisovic, {RO-Crate Community}, Groth, and Goble]{soilandreyes2022rocrate}
Stian Soiland-Reyes, Peter Sefton, Merc{\`e} Crosas, Leyla~Jael Castro, Frederik Coppens, Jos{\'e}~M. Fern{\'a}ndez, Daniel Garijo, Bj{\"o}rn Gr{\"u}ning, Marco La~Rosa, Simone Leo, Eoghan {\'O}~Carrag{\'a}in, Marc Portier, Ana Trisovic, {RO-Crate Community}, Paul Groth, and Carole Goble.
\newblock Packaging research artefacts with {RO-Crate}.
\newblock \emph{Data Science}, 5\penalty0 (2):\penalty0 97--138, 2022.
\newblock \doi{10.3233/DS-210053}.

\bibitem[Leo et~al.(2024)Leo, Crusoe, Rodr{\'i}guez-Navas, Sirvent, Kanitz, et~al.]{leo2024wrroc}
Simone Leo, Michael~R. Crusoe, Laura Rodr{\'i}guez-Navas, Ra{\"u}l Sirvent, Alexander Kanitz, et~al.
\newblock Recording provenance of workflow runs with {RO-Crate}.
\newblock \emph{PLOS ONE}, 19\penalty0 (9):\penalty0 e0309210, 2024.
\newblock \doi{10.1371/journal.pone.0309210}.

\bibitem[Abolhasani and Kumacheva(2023)]{abolhasani2023selfdriving}
Milad Abolhasani and Eugenia Kumacheva.
\newblock The rise of self-driving labs in chemical and materials sciences.
\newblock \emph{Nature Synthesis}, 2\penalty0 (6):\penalty0 483--492, 2023.
\newblock \doi{10.1038/s44160-022-00231-0}.

\bibitem[Szymanski et~al.(2023)Szymanski, Rendy, Fei, Kumar, He, Milsted, et~al.]{szymanski2023alab}
Nathan~J. Szymanski, Bernardus Rendy, Yuxing Fei, Rishi~E. Kumar, Tanjin He, David Milsted, et~al.
\newblock An autonomous laboratory for the accelerated synthesis of inorganic materials.
\newblock \emph{Nature}, 624\penalty0 (7990):\penalty0 86--91, 2023.
\newblock \doi{10.1038/s41586-023-06734-w}.

\bibitem[Honarmandi et~al.(2019)Honarmandi, Duong, Ghoreishi, Allaire, and Arroyave]{honarmandi2019bayesian}
Pejman Honarmandi, Thien~Chi Duong, S~Fatheme Ghoreishi, Douglas Allaire, and Raymundo Arroyave.
\newblock Bayesian uncertainty quantification and information fusion in calphad-based thermodynamic modeling.
\newblock \emph{Acta Materialia}, 164:\penalty0 636--647, 2019.

\bibitem[Zhu et~al.(2025{\natexlab{a}})Zhu, Sar{\i}t{\"u}rk, and Arr{\'o}yave]{zhu2025accelerating}
Siya Zhu, Do{\u{g}}uhan Sar{\i}t{\"u}rk, and Raymundo Arr{\'o}yave.
\newblock Accelerating {CALPHAD}-based phase diagram predictions in complex alloys using universal machine learning potentials: Opportunities and challenges.
\newblock \emph{Acta Materialia}, page 120747, 2025{\natexlab{a}}.

\bibitem[Zhu et~al.(2025{\natexlab{b}})Zhu, Sar{\i}t{\"u}rk, and Arr{\'o}yave]{zhu2025machinelearningpotentialsalloys}
Siya Zhu, Do{\u{g}}uhan Sar{\i}t{\"u}rk, and Raymundo Arr{\'o}yave.
\newblock Machine learning potentials for alloys: A detailed workflow to predict phase diagrams and benchmark accuracy, 2025{\natexlab{b}}.
\newblock arXiv:2506.16771.

\bibitem[Kunselman et~al.(2025{\natexlab{a}})Kunselman, Zhu, Sar{\i}t{\"u}rk, and Arr{\'o}yave]{kunselman2025ptw}
Courtney Kunselman, Siya Zhu, Do{\u{g}}uhan Sar{\i}t{\"u}rk, and Raymundo Arr{\'o}yave.
\newblock Construction and tuning of {CALPHAD} models using machine-learned interatomic potentials and experimental data: A case study of the {Pt--W} system.
\newblock \emph{Journal of Phase Equilibria and Diffusion}, 2025{\natexlab{a}}.
\newblock \doi{10.1007/s11669-025-01222-2}.

\bibitem[Kunselman et~al.(2024)Kunselman, Bocklund, van~de Walle, Otis, and Arr{\'o}yave]{kunselman2024analytically}
Courtney Kunselman, Brandon Bocklund, Axel van~de Walle, Richard Otis, and Raymundo Arr{\'o}yave.
\newblock Analytically differentiable metrics for phase stability.
\newblock \emph{Calphad}, 86:\penalty0 102705, 2024.

\bibitem[Kunselman et~al.(2025{\natexlab{b}})Kunselman, Bocklund, Otis, and Arroyave]{kunselman2025analytical}
Courtney Kunselman, Brandon Bocklund, Richard Otis, and Raymundo Arroyave.
\newblock Analytical gradient-based optimization of {CALPHAD} model parameters.
\newblock \emph{arXiv preprint arXiv:2505.01585}, 2025{\natexlab{b}}.
\newblock \doi{10.48550/arXiv.2505.01585}.
\newblock URL \url{https://arxiv.org/abs/2505.01585}.
\newblock arXiv:2505.01585 [cond-mat.mtrl-sci].

\bibitem[Sar{\i}t{\"u}rk et~al.(2025)Sar{\i}t{\"u}rk, Kunselman, and Zhu]{sariturk2025phaseforgeplus}
Do{\u{g}}uhan Sar{\i}t{\"u}rk, Courtney Kunselman, and Siya Zhu.
\newblock {PhaseForgePlus}, 2025.
\newblock Software, version 0.1.0.

\bibitem[{Dictionary.com}(2023)]{dictionarycom_sage}
{Dictionary.com}.
\newblock sage (noun).
\newblock Dictionary.com Unabridged, based on the Random House Unabridged Dictionary, \url{https://www.dictionary.com/browse/sage}, 2023.
\newblock Accessed 11 November 2026.

\bibitem[{Anthropic}(2025)]{mcp}
{Anthropic}.
\newblock Model context protocol specification.
\newblock Version 2025-06-18, 2025.
\newblock URL \url{https://modelcontextprotocol.io/specification/2025-06-18}.

\bibitem[Lee(1992)]{lee1992}
Byeong-Joo Lee.
\newblock On the stability of {Cr} carbides.
\newblock \emph{Calphad}, 16\penalty0 (2):\penalty0 121--149, 1992.
\newblock \doi{10.1016/0364-5916(92)90002-f}.

\bibitem[Hao et~al.(2021)Hao, Ruban, and Xiong]{hao2021}
Liangyan Hao, Andrei Ruban, and Wei Xiong.
\newblock {CALPHAD} modeling based on {Gibbs} energy functions from zero kevin and improved magnetic model: A case study on the {Cr--Ni} system.
\newblock \emph{Calphad}, 73:\penalty0 102268, 2021.
\newblock \doi{10.1016/j.calphad.2021.102268}.

\bibitem[Kunselman et~al.(2026)Kunselman, Bocklund, Otis, and Arr{\'o}yave]{kunselman2026hmc}
Courtney Kunselman, Brandon Bocklund, Richard Otis, and Raymundo Arr{\'o}yave.
\newblock Gradient-informed {H}amiltonian {M}onte {C}arlo for multicomponent {CALPHAD} model optimization and uncertainty quantification.
\newblock \emph{Acta Materialia}, 315:\penalty0 122412, 2026.
\newblock \doi{10.1016/j.actamat.2026.122412}.

\bibitem[Hong and Liu(2025)]{hong2025entropy}
Qi-Jun Hong and Zi-Kui Liu.
\newblock Generalized approach for rapid entropy calculation of liquids and solids.
\newblock \emph{Physical Review Research}, 7:\penalty0 L012030, 2025.
\newblock \doi{10.1103/PhysRevResearch.7.L012030}.

\bibitem[{Schema.org}(2026)]{schemaorg_howto}
{Schema.org}.
\newblock {HowTo}, {HowToSection} and {HowToStep} types.
\newblock Schema.org vocabulary, 2026.
\newblock URL \url{https://schema.org/HowTo}.
\newblock Accessed 7 November 2026.

\bibitem[Pugliese and Fitterer(1970)]{pugliese1970}
L.~A. Pugliese and G.~R. Fitterer.
\newblock Activities and phase boundaries in the {Cr-Ni} system using a solid electrolyte technique.
\newblock \emph{Metallurgical Transactions}, 1:\penalty0 1997--2002, 1970.
\newblock \doi{10.1007/BF02642800}.

\bibitem[Dinsdale(1991)]{dinsdale1991sgte}
A.~T. Dinsdale.
\newblock {SGTE} data for pure elements.
\newblock \emph{Calphad}, 15\penalty0 (4):\penalty0 317--425, 1991.
\newblock \doi{10.1016/0364-5916(91)90030-N}.

\bibitem[Zunger et~al.(1990)Zunger, Wei, Ferreira, and Bernard]{zunger1990special}
Alex Zunger, S-H Wei, Luiz~G Ferreira, and James~E Bernard.
\newblock Special quasirandom structures.
\newblock \emph{Physical review letters}, 65\penalty0 (3):\penalty0 353, 1990.

\bibitem[Arr{\'o}yave(2026{\natexlab{a}})]{arroyave2026mscraps}
Raymundo Arr{\'o}yave.
\newblock m\_scraps: Multi-sublattice short-range-order and cluster-correlation targeting, 2026{\natexlab{a}}.
\newblock Software, version 0.6.0.

\bibitem[Singh et~al.(2021)Singh, Sharma, Singh, Balasubramanian, and Johnson]{singh2021scraps}
Rahul Singh, Aayush Sharma, Prashant Singh, Ganesh Balasubramanian, and Duane~D. Johnson.
\newblock Accelerating computational modeling and design of high-entropy alloys.
\newblock \emph{Nature Computational Science}, 1:\penalty0 54--61, 2021.
\newblock \doi{10.1038/s43588-020-00006-7}.

\bibitem[Bochkarev et~al.(2024)Bochkarev, Lysogorskiy, and Drautz]{bochkarev2024grace}
Anton Bochkarev, Yury Lysogorskiy, and Ralf Drautz.
\newblock Graph atomic cluster expansion for semilocal interactions beyond equivariant message passing.
\newblock \emph{Physical Review X}, 14:\penalty0 021036, 2024.
\newblock \doi{10.1103/PhysRevX.14.021036}.

\bibitem[Liu et~al.(2016)Liu, Lindwall, Gheno, and Liu]{liu2016alcocrni}
Xuan~L. Liu, Greta Lindwall, Thomas Gheno, and Zi-Kui Liu.
\newblock Thermodynamic modeling of {Al--Co--Cr}, {Al--Co--Ni}, {Co--Cr--Ni} ternary systems towards a description for {Al--Co--Cr--Ni}.
\newblock \emph{Calphad}, 52:\penalty0 125--142, 2016.
\newblock \doi{10.1016/j.calphad.2015.12.007}.

\bibitem[Bracq et~al.(2021)Bracq, Crivello, Laurent-Brocq, Joubert, and Guillot]{bracq2021enthalpy}
Guillaume Bracq, Jean-Claude Crivello, Mathilde Laurent-Brocq, Jean-Marc Joubert, and Ivan Guillot.
\newblock What is the enthalpy contribution to the stabilization of the {Co--Cr--Fe--Mn--Ni} face-centered cubic solid solution?
\newblock \emph{Journal of Phase Equilibria and Diffusion}, 42:\penalty0 561--570, 2021.
\newblock \doi{10.1007/s11669-021-00922-9}.

\bibitem[Kirklin et~al.(2015)Kirklin, Saal, Meredig, Thompson, Doak, Aykol, R{\"u}hl, and Wolverton]{kirklin2015oqmd}
Scott Kirklin, James~E. Saal, Bryce Meredig, Alex Thompson, Jeff~W. Doak, Muratahan Aykol, Stephan R{\"u}hl, and Chris Wolverton.
\newblock The {Open Quantum Materials Database} ({OQMD}): Assessing the accuracy of {DFT} formation energies.
\newblock \emph{npj Computational Materials}, 1:\penalty0 15010, 2015.
\newblock \doi{10.1038/npjcompumats.2015.10}.

\bibitem[Wales and Doye(1997)]{wales1997basinhopping}
David~J. Wales and Jonathan P.~K. Doye.
\newblock Global optimization by basin-hopping and the lowest energy structures of {Lennard-Jones} clusters containing up to 110 atoms.
\newblock \emph{The Journal of Physical Chemistry A}, 101\penalty0 (28):\penalty0 5111--5116, 1997.
\newblock \doi{10.1021/jp970984n}.

\bibitem[Xiang et~al.(1997)Xiang, Sun, Fan, and Gong]{xiang1997dualannealing}
Y.~Xiang, D.~Y. Sun, W.~Fan, and X.~G. Gong.
\newblock Generalized simulated annealing algorithm and its application to the {Thomson} model.
\newblock \emph{Physics Letters A}, 233\penalty0 (3):\penalty0 216--220, 1997.
\newblock \doi{10.1016/S0375-9601(97)00474-X}.

\bibitem[Virtanen et~al.(2020)Virtanen, Gommers, Oliphant, Haberland, Reddy, et~al.]{virtanen2020scipy}
Pauli Virtanen, Ralf Gommers, Travis~E. Oliphant, Matt Haberland, Tyler Reddy, et~al.
\newblock {SciPy} 1.0: Fundamental algorithms for scientific computing in {Python}.
\newblock \emph{Nature Methods}, 17\penalty0 (3):\penalty0 261--272, 2020.
\newblock \doi{10.1038/s41592-019-0686-2}.

\bibitem[Cahn(1961)]{cahn1961spinodal}
John~W. Cahn.
\newblock On spinodal decomposition.
\newblock \emph{Acta Metallurgica}, 9\penalty0 (9):\penalty0 795--801, 1961.
\newblock \doi{10.1016/0001-6160(61)90182-1}.

\bibitem[Tierney and Kadane(1986)]{tierney1986laplace}
Luke Tierney and Joseph~B. Kadane.
\newblock Accurate approximations for posterior moments and marginal densities.
\newblock \emph{Journal of the American Statistical Association}, 81\penalty0 (393):\penalty0 82--86, 1986.
\newblock \doi{10.1080/01621459.1986.10478240}.

\bibitem[Svechnikov and Pan(1962)]{svechnikov1962characteristics}
V.~N. Svechnikov and V.~M. Pan.
\newblock Characteristics of the equilibrium diagram and processes of solution and precipitation in the {Cr--Ni} system.
\newblock \emph{Sbornik Nauchnykh Rabot Instituta Metallofiziki Akademii Nauk Ukrainskoi SSR}, 15:\penalty0 164--178, 1962.

\bibitem[Dench(1963)]{dench1963adiabatic}
W.~A. Dench.
\newblock Adiabatic high-temperature calorimeter for the measurement of heats of alloying.
\newblock \emph{Transactions of the Faraday Society}, 59:\penalty0 1279--1292, 1963.
\newblock \doi{10.1039/TF9635901279}.

\bibitem[Collins(1988)]{collins1988electron}
M.~J. Collins.
\newblock Electron optic determination of solid phase boundaries in {Ni--Cr} system.
\newblock \emph{Materials Science and Technology}, 4\penalty0 (6):\penalty0 560--561, 1988.
\newblock \doi{10.1179/mst.1988.4.6.560}.

\bibitem[Jette et~al.(1934)Jette, Nordstrom, Queneau, and Foote]{jette1934xray}
E.~R. Jette, V.~H. Nordstrom, B.~Queneau, and F.~Foote.
\newblock X-ray studies on the nickel-chromium system.
\newblock \emph{Transactions of the AIME}, 111:\penalty0 361--373, 1934.

\bibitem[Guggenheim(1944)]{guggenheim1944mixtures}
E.~A. Guggenheim.
\newblock Statistical thermodynamics of mixtures with zero energies of mixing.
\newblock \emph{Proceedings of the Royal Society of London. Series A}, 183:\penalty0 203--212, 1944.
\newblock \doi{10.1098/rspa.1944.0032}.

\bibitem[Kikuchi(1951)]{kikuchi1951cooperative}
Ryoichi Kikuchi.
\newblock A theory of cooperative phenomena.
\newblock \emph{Physical Review}, 81:\penalty0 988--1003, 1951.
\newblock \doi{10.1103/PhysRev.81.988}.

\bibitem[Arr{\'o}yave(2026{\natexlab{b}})]{pairentropy}
Raymundo Arr{\'o}yave.
\newblock pair-entropy: Pair (quasi-chemical) estimator of the excess entropy of liquids from molecular dynamics.
\newblock \url{https://github.com/ArroyaveLab/pair-entropy}, 2026{\natexlab{b}}.

\end{thebibliography}

\end{document}